\documentclass[runningheads,a4paper]{llncs}

\usepackage{amssymb}
\usepackage{graphicx}

\usepackage{url}
\urldef{\mailsa}\path|{jan, iweber}@qf.org.qa|
\urldef{\mailsb}\path|anna.kramer, leonie.kunz, christine.reiss, nicole.sator,|
\urldef{\mailsc}\path|erika.siebert-cole, peter.strasser, lncs}@springer.com|    
\newcommand{\keywords}[1]{\par\addvspace\baselineskip
\noindent\keywordname\enspace\ignorespaces#1}
\usepackage{booktabs}
\usepackage{dcolumn}

\usepackage{balance}  % to better equalize the last page
\usepackage{graphics} % for EPS, load graphicx instead
\usepackage{times}    % comment if you want LaTeX's default font
\usepackage{url}      % llt: nicely formatted URLs
\usepackage{amsmath}
\usepackage[usenames,dvipsnames]{xcolor}
\usepackage{subfigure}
\usepackage{multicol}

\usepackage[utf8]{inputenc} %unicode support
\usepackage{times}
\usepackage{helvet}
\usepackage{courier}

\usepackage{xcolor}
\usepackage{graphicx}
\usepackage{epstopdf}
\usepackage{amssymb}% http://ctan.org/pkg/amssymb
\usepackage{pifont}% http://ctan.org/pkg/pifont
\usepackage{enumitem}

\newcounter{NumberOfComments}
\stepcounter{NumberOfComments}

\definecolor{DarkGreen}{rgb}{0.000000,0.6,0.000000 } 

\newcounter{JSNumberOfComments}
\stepcounter{JSNumberOfComments}

\begin{document}

\mainmatter  % start of an individual contribution

\title{Whom Should We Sense in ``Social Sensing'' -- Analyzing Which Users Work Best for Social Media Now-Casting}

\titlerunning{Whom Should We Sense in ``Social Sensing''}

% \author{Alfred Hofmann%
% \thanks{Please note that the LNCS Editorial assumes that all authors have used
% the western naming convention, with given names preceding surnames. This determines
% the structure of the names in the running heads and the author index.}%
% \and Ursula Barth\and Ingrid Haas\and Frank Holzwarth\and\\
% Anna Kramer\and Leonie Kunz\and Christine Rei\ss\and\\
% Nicole Sator\and Erika Siebert-Cole\and Peter Stra\ss er}
%
\author{Jisun An\and Ingmar Weber}
% \authorrunning{Lecture Notes in Computer Science: Authors' Instructions}

\institute{Qatar Computing Research Institute, Hamad bin Khalifa University, Doha, Qatar\\
\mailsa}
% \mailsb\\
% \mailsc\\
% \url{http://www.qcri.com/our-research/social-computing}}


\toctitle{Lecture Notes in Computer Science}
\tocauthor{Authors' Instructions}
\maketitle

\begin{abstract}
Given the ever increasing amount of publicly available social media data, there is growing interest in using online data to study and quantify phenomena in the offline ``real'' world. As social media data can be obtained in near real-time and at low cost, it is often used for ``now-casting'' indices such as levels of flu activity or unemployment. The term ``social sensing'' is often used in this context  to describe the idea that users act as ``sensors'', publicly reporting their health status or job losses. Sensor activity during a time period is then typically aggregated in a ``one tweet, one vote'' fashion by simply counting. At the same time, researchers readily admit that social media users are not a perfect representation of the actual population. Additionally, users differ in the amount of details of their personal lives that they reveal. Intuitively, it should be possible to improve now-casting by assigning different weights to different user groups.
 
In this paper,  we ask ``How does social sensing actually work?'' or, more precisely, ``Whom should we sense--and whom not--for optimal results?''. We investigate how different sampling strategies affect the performance of now-casting of two common offline indices: flu activity and unemployment rate. We show that now-casting can be improved by 1) applying user filtering techniques and 2) selecting users with complete profiles. We also find that, using the right type of user groups, now-casting performance does not degrade, even when drastically reducing the size of the dataset. More fundamentally, we describe which type of users contribute most to the accuracy by asking if ``babblers are better''. We conclude the paper by providing guidance on how to select better user groups for more accurate now-casting.
\end{abstract}

\footnotetext[1]{This is a pre-print of a forthcoming EPJ Data Science paper. }

%%%%%%%%%%%%%%%%%%%%%%%%% start of article main body
% <put your article body there>

%%%%%%%%%%%%%%%%
%% Introduction %%
%%
\section{Introduction}
There is a growing amount of interest in using online social media data to study phenomena in the offline ``real'' world. Applications range from flu tracking and epidemiology, to now-casting unemployment and other economic indicators, to election prediction and public opinion monitoring. Often, the term ``social sensing'' is used to describe the idea that normal social media users act as ``sensors'', reporting their health status, job losses and voting intentions. Following this idea, such methods should work best if there is a large number of normal people on social media who report on all minutiae of their daily lives.
 
However, there is no logical necessity for now-casting to work best in such settings. For example, if you wanted to now-cast today's movie box office sales then it might be best to analyze tweets from cinemas on Twitter. Cinemas are likely tweeting their program, and the number of cinemas showing a particular movie could be a better estimate for today's box office sales than a noisier estimate derived from ``normal'' tweets. On the other hand, it is probably not desirable to only monitor users who are always tweeting about their personal health when it comes to monitoring flu epidemics. These constantly self-diagnosing users are likely to follow a different disease cycle than the general population, leading to sub-optimal estimates. In other domains, such as predicting tomorrow's stock price from Twitter activity, it might be desirable to only take experts into account and to ignore normal people altogether.
 
Despite this fundamental question -- whom \emph{should} we sense when we ``social sense'' -- the usual approach is ``one tweet, one vote'' where all tweets are treated equally and only spam is removed. This paper goes beyond such a simplistic methodology and explores which users groups are most desirable to include. For example, is it good to have a lot of ``unfiltered'' users in the data set, who share lots of details about their private lives, or is it preferable to have more ``reserved'' users, who typically do not tweet about daily life but who, once in a while, report on being sick? Related to this, are there particular demographic groups who should be given more or less attention, potentially because they are underrepresented?
 
We focus our analysis on two application domains, now-casting of flu activity and unemployment. We choose these domains as (i) they come with comparably little ``astro-turfing'' and (ii) they have hard, objective ``ground truth''. Tracking public opinions with, say, the goal of predicting election outcomes comes with additional challenges to detect political campaigns or even to establish reliable ground truth time series of ``opinions''. Even the two domains we chose are expected to show different characteristics as, e.g., it is less of a stigma to tweet ``I'm down with the flu'' than it is to tweet ``I've just lost my job''.

For our evaluation, we interpret ``social sensing'' as ``social media based now-casting''. Though there are other applications, such as event detection, we feel that now-casting represents well the idea of using individuals as sensors. We structure our analysis around the following research hypotheses, elaborated on in Section~\ref{sec:user_group_now_casting}:
\begin{itemize}
\item[] $[H1]$ A more complete Twitter profile is a better sensor.
\item[] $[H2]$ A user's Twitter stats are as good as their demographics in predicting their value as a social sensor.
\item[] $[H3]$ Babblers are best and users who share many personal details are great social sensors.
\item[] $[H4]$ Better data beats bigger data and no full ``Firehose'' is needed.
\item[] $[H5]$ Giving more weight to underrepresented regions helps.
\item[] $[H6]$ Giving more weight to barely active users increases the now-casting performance.
\end{itemize}

This paper differs from previous work on now-casting in that our goal is not to report a certain correlation between online and offline indices. Rather, we offer a systematic attempt at \emph{explaining} which user types are desirable to monitor as social sensors, and which are not. We believe that the insights gained from investigating the hypotheses above and from an in-depth discussion (Section~\ref{sec:discussion}) are relevant to a wide range of applications. In particular, we provide general guidelines as to when social media based now-casting is expected to work, and when not.

%%%%%%%%%%%%%%%%
%% Background %%
%%
\section{Related Work}

\subsection{Social Sensing}
The term ``social sensing'' is used differently by different communities. In the area of mobile computing, it usually refers to providing users with physical, wearable sensors that detect a user's environment, often including social interactions and being close to other sensors~\cite{OlguinP08,MadanCLP10,AliSSONM11,LiuYLF11,AggarwalA13}. This type of sensing-through-physical-devices is not how we use the term in this paper.

For the purpose of this work, we define social sensing as \emph{using public social media data to make statements about the ``real'', offline world}. An example application is event detection~\cite{SakakiOM10,CepniA14,AlbakourMO13}. A more ``standardized'' application of social sensing is \emph{now-casting} where the goal is to use social media data to predict the next or current element in a time series representing some offline activity. This application scenario is what we are using for this paper and we will next survey related work.

\subsection{Social Media Now-casting}
When it comes to now-casting, the poster child usually mentioned is Google Flu Trends~\cite{ginsberg2008detecting}, though limitations have been pointed out~\cite{lazer2014parable}. Web search volume has also been used for a number of other now-casting scenarios, including stock prices and volume~\cite{preis2012quantifying,bordino2012web}. However, social media is also being used more and more for such tasks and we briefly review some existing work.

Monitoring populations for flu epidemics has gained the attention of social media researchers~\cite{SzomszorKQ10,Culotta13,LamposBC10,AramakiMM11}. We use flu now-casting as one of the test cases in our experiments. 
One of the most popular applications of social media now-casting is public opinion monitoring and election prediction~\cite{connor2010@icwsm,TumasjanSSW10}. However, early claims of success in this area have come under scrutiny and the feasibility of such an endeavor seems uncertain~\cite{Gayo-Avello12,MetaxasMG11,Bravo-MarquezGMP12}. In our experiments, we do not use public opinion or election data and we discuss how the inclusion of such data might affect the results in the Discussion section.
Socio-economic variables, including consumer confidence, have also been used in now-casting studies~\cite{bollen2011modeling,connor2010@icwsm}. We use a similar setup as in~\cite{antenucci2014} for now-casting unemployment data, but our focus is on testing the effect of including or excluding different user groups from the analysis. 

\subsection{Twitter Representativeness and Bias}
Though the lack of representativeness of Twitter data is widely acknowledged in studies using this data, there are few studies that have systematically studied bias related to this data. One important study in this regard is~\cite{amislove2011@icwsm} who study differences in the distribution between offline census data and Twitter users for gender, geography and race. They observed a male-dominated Twitter population, concentrated in urban areas, with geographic patterns varying as to which race is over- or underrepresented. Since this study was done Twitter's user base has undergone a considerable change and, e.g., the male dominance might no longer be true~\cite{liu2014tweets}. 

Apart from the bias linked to who is using Twitter, there is also a technical bias imposed by how the data is collected. Though the Streaming API is usually assumed to give a uniform sample of tweet activity, this was found not to be the case~\cite{morstatter2013sample}. Similarly, researchers have observed that depending on whether the streaming or search APIs are used different biases in the reconstruction of mention and retweet networks arise~\cite{gonzalez2014assessing}.

Even though bias does exist in Twitter data, it is still possible to extract meaningful signals. In their recent work, Zagheni and Weber provide a general framework to extract information from biased web data by using ``difference-in-differences'' as one proposed methods. This approach does not try to estimate \emph{absolute} levels of real-world variables but, rather, it aims to obtain reliable estimates of \emph{trends}~\cite{Zagheni2015}.

\subsection{Inferring Demographic Information}

Some of our filtering schemes make use of demographic information such as age or gender. Unlike Facebook or other social media, Twitter does not provide a structured field where users can enter this information. As knowing basic demographic information about users adds a lot of value to almost any study, a lot of efforts have been taken to infer these variables. Methods differ wildly in how much information they require about individual users and, generally, having \emph{network} information makes the task easier as the classification label of a set of neighboring nodes carries a lot of informational content~\cite{ZamalLR12}.

Gender is usually considered one of the easier demographic dimensions to infer. The most basic approach is to use a gender-based dictionary, often based on census data~\cite{liu2013,amislove2011@icwsm} and there are web services that can be used for this purpose\footnote{\url{http://genderize.io/}}. Tweet content has also been used, in particular for non-English languages where the form of adjectives can often reveal the gender of the speaker~\cite{cohen2013}. Even the profile background color has been observed to provide clues on the gender of the user~\cite{AlowibdiBY13}. Lastly, the gender of a user can also be inferred using the profile picture along with image processing, such as provided by Face++\footnote{\url{http://www.faceplusplus.com/demo-detect/}}, which is useful for studies across languages and naming conventions~\cite{ZagheniGWS14}. For our study, we used the gender dictionary used in~\cite{magnoweber14socinfo}, which combines existing first-name dictionaries with a new dictionary derived from a large Google+ dataset.

Age is harder to infer but another important variable to know. Tweet content would be the typical feature set of choice for this task~\cite{nguyen2013@icwsm}. In our study, where we do not have enough longitudinal tweet data for users to infer their age, we made use of the Face++ service to infer approximate age from the profile pictures of users.

Geographic location is another useful variable to have, also because it is linked to other variables such as income or race if the geography can be inferred with sufficient resolution. There is a large body of academic work on this problem, typically using longitudinal data and looking for place-specific references~\cite{ChengCL10,MahmudND12,IkawaET12}, but using social network information has also proved to be useful~\cite{DavisPOA11,PontesMV0AKA12}. In our study, we limited ourselves to inferring state-level locations and used the user-provided location fields in combination with a dictionary previously used in~\cite{ChenWO14}.

There are many other potential variables that could be inferred, but which we did not use for this study. These include political orientation~\cite{cohen2013,PennacchiottiP11,PennacchiottiP11b}, religious affiliation~\cite{NguyenL14,ChenWO14}, or ethnicity and race~\cite{amislove2011@icwsm,PennacchiottiP11,PennacchiottiP11b}.

\section{Research Hypotheses}
\label{sect:hypotheses}

We frame our study around a set of research hypotheses. These hypotheses summarize how one might intuitively suspect social sensing to work. \\

\noindent \textbf{More Complete Profile = Better Profile}.
\textit{[H1] Users with a more complete profile (specified location, profile picture with recognizable face, ...) provide better social sensing data.}

The ``Twitter Egg'' is the default picture of a newly created Twitter account. Knowing that a user has invested the effort to replace this picture by a profile picture could be taken as indication of both engagement and diligence. Similarly, if the user provides a valid location, rather than the empty default or a non--revealing ``at home'', then this could indicate that the user is less concerned about privacy, more open in sharing personal details, and  thus a more reliable social sensor. We hypothesize that, indeed, users with a more complete profile provide better data for now-casting. \\

\noindent \textbf{Twitter Stats Better Filter Than Demographics}. \textit{[H2] Twitter statistics, such as a user's number of followers or tweets, are at least as useful as demographic information in determining good social sensors.}

Obtaining all of a user's historic tweets comes with technical challenges related to Twitter API restrictions and bandwidth constraints. Thus, one would like to predict a user's quality as a social sensor based solely on information that can be (noisily) derived from their profile, which includes their name and their profile picture. For example, a user's gender could be predictive of how they use Twitter and, hence, how valuable their tweets are for social sensing. At the same time, Twitter-intrinsic ``demographics'' such as the age of the account or the number of followers possibly gives a better signal for how the user uses Twitter. We hypothesize that a user's Twitter statistics are better predictors for their value in now-casting than inferred demographics such as gender. \\

\noindent \textbf{Babblers are Best}.
\textit{[H3] Users who tweets about their daily lives are better social sensors than those mostly discussing professional or public topics.}

Intuitively, one would expect the most ``honest'' signals to come from users who are most open about sharing anything on social media. Given that someone shares details about their food consumption, the movies they watch, their day at the office and so on, we might also expect them to truthfully report on being sick for example. We hypothesize that users with a significant fraction of tweets about their private, daily lives provide better data for now-casting than more reserved users. \\

\noindent \textbf{Better Data Beats Bigger Data}.
\textit{[H4] One can obtain equal or better now-casting results by using a drastically reduced dataset, as long as the data quality of the smaller set is high.}

A sample size of n=1,000 would be typical for a high--quality survey. What surveys lack in size, compared to n=1,000,000 in many social media studies, they make up in data quality. We hypothesize that with the right kind of data filtering approach, the vast majority of data can be ``thrown away'' without degrading now-casting performance. \\

\noindent \textbf{Geographic Reweighting : Putting Weight Where it Belongs}.

\textit{[H5] By giving more weight to geographical segments of the population that are under--represented on Twitter the now-casting performance increases.}

Twitter penetration rates vary across geographic regions. Whereas in urban areas typically a larger fraction of the population is active on social media, such technologies are less common in rural areas. This means that merely summing counts across a country tends to put too much emphasis on urban centers. Given estimates of the Twitter penetration rates and information about a user's location, this can, however, be corrected for by giving more weight to underrepresented areas.\\

\noindent \textbf{Up-Weighting Inactive Users: Boosting the Silent Majority}. 
\textit{[H6] Giving more weight to inactive users--those who tweet less--increases the now-casting performance}.

The majority of users tend to tweet less than once per day and only about 20\% of monthly active users are also daily active users\footnote{\url{http://blog.peerreach.com/2013/11/4-ways-how-twitter-can-keep-growing/}}, while 40\% of users do not tweet at all but only watch other people tweet\footnote{\url{http://www.statisticbrain.com/twitter-statistics/}}. So we hypothesize that when they do mention something, it is more significant. Also inactive users represent a larger user base and their voice should hence be ``amplified''.

\section{Data Collection}
\label{sec:data_collection}

To evaluate the value of different user groups for social sensing, we focus our interest on the following now-casting tasks: flu activity and unemployment rate. We select these tasks due to a combination of (i) reasonably high frequency changes -- without any changes there is nothing to track, (ii) fairly large amounts of data available via Twitter, and (iii) availability of ``hard'' offline data, also see the discussion at the end. We note that both offline and Twitter datasets are collected for a period between January and November 2014.

\subsection{Offline indices}

\noindent \textbf{\textit{Flu activity}}. 
The U.S.\ Centers for Disease Control and Prevention (CDC) publishes weekly reports from the U.S.\ Outpatient Influenza-like Illness Surveillance Network (ILINet). ILINet monitors over 3,000 health providers nationwide to report the fraction of patients with influenza-like illnesses (ILI). The aggregated numbers for each HHS region\footnote{\url{http://www.hhs.gov/about/regionmap.html}} are publicly available\footnote{\url{http://gis.cdc.gov/grasp/fluview/fluportaldashboard.html}}. 
While ILINet is a valuable tool in detecting influenza outbreaks, it suffers from a high operational cost and slow reporting time, typically a one to two week delay. Concerning data size, ILINet reported 565,134 cases with flu symptoms from January to November 2014. Aggregated at the level of a week, there are 12,024 cases on average with a minimum of 4,729, a median of 10,817, and a maximum of 28,721.

\noindent \textbf{\textit{Unemployment Rate}}. We collect Unemployment Insurance (UI) Weekly Claims Data\footnote{\url{http://workforcesecurity.doleta.gov/unemploy/claims.asp}} from the U.S.\ Department of Labor. The initial claims data are well-suited for our study. For the 11 months of our study, the total number of UI claims is 13,434,657. When aggregated weekly, the minimum is 227,571, the median is 288,748, the mean is 298,548 and the maximum is 534,966.

\begin{table*}[h!]
\caption{\textbf{Summary of our Twitter dataset. The keywords used for extracting unemployment tweets were ``got fired'', ``lost ** job'', ``get a job'' and ``unemployment''. For flu tweets only ``flu'' was used. Both sets of keywords were used in previous work. The value in parentheses is the fraction of users with the corresponding inferred demographics.}}
    \begin{center}
    % \vspace*{-5mm}
    \small \frenchspacing
    \begin{tabular}{c|cc|cc}
    \hline
    \hline
Topic & Flu Activity & Unemployment & 1st-person Flu & 1st-person Unemp. \\
\hline
\#Tweets & 153,848 & 145,780 &  79,223 & 83,015 \\
\#Authors & 142,458 & 139,300 & 75,000 & 72,375 \\
State & 56,967 (40\%) & 48,670 (35\%) & 24,287 (32\%) & 22,987 (32\%) \\
Gender & 56,903 (40\%)  &  49,359 (36\%) & 28,187 (38\%)  & 23,555 (33\%) \\
Male & 24,896 (18\%) & 29,018 (20\%) & 10,664 (14\%) &11,911 (17\%) \\
Female & 32,009 (23\%) & 22,146 (15\%) & 17,524 (23\%) &11,644 (16\%)\\
Age & 42,916  (30\%) & 42,049 (30\%) &23,418 (31\%) &22,584 (31\%)\\
Age $<$ 24 & 19,682 (14\%)&20,452 (15\%) & 12,745 (17\%)& 12,419 (17\%)\\
Age $>=$ 24 &23,234 (16.3\%) &  21,597 (16\%)&10,673 (14\%)&10,165 (14\%)\\
\hline
\hline
    \end{tabular}
    \end{center}
    % \vspace*{-5mm}
    \label{tab:summary}
\end{table*}

\subsection{Twitter datasets}

For this study, we obtained access to historic ``Decahose'' Twitter data. This represents a 10\% sample of all public tweets for the time period 2014/01/01--2014/11/30. From this data, we extract two different datasets corresponding to the two topics of interest by collecting tweets using keyword matching. 
For the flu dataset, we collect tweets containing the keyword `flu'. For our unemployment dataset, we use a list of keywords: axed, canned, downsized, pink slip, get a job, got fired, lost ** job, laid off, and unemployment, as suggested by~\cite{antenucci2014}.

The initial Twitter collection includes 222,527 tweets for flu and 250,092 for unemployment. Since we are interested in finding real activity of individuals, we removed all retweets, that is tweets with ``RT @''. Then, as a basic spam removal, we keep only those users whose language is specified as ``en'', the self-description bio is not empty, their tweet count is greater than 10 and less than 50K, their follower count is at least 10 and the count of days since joining Twitter is at least 10, leaving us with 153,848 tweets for flu and 145,780 for unemployment. This process resulted in a collection of 299,628 tweets with 256,154 users. Table~\ref{tab:summary} shows a summary of our datasets. 

One caveat of simple keyword matching is that it includes false positives. For example, Culotta showed that excluding terms `h1n1' and `swine flu' improves the fit between Twitter and offline flu data~\cite{Culotta13}. Since the period of our data collection is different from the one used in~\cite{Culotta13}, the terms triggering false alarms are likely to be different. Thus, we build a classifier that extracts ``first person'' tweets for each of the two topics. This helps to reduce the effect of, say, agencies tweeting about flu to promote vaccination campaigns and is described in the following.

\subsection{First person tweet classifier}
\label{sec:firstperson_tweet_classifier}
A tweet with a keyword such as ``flu'' can be an act of social sensing, where an individual reports flu symptoms. But it can also be about a news article relating to the flu season. Intuitively, ``first person reports'' should be better indicators for now-casting and we build a classifier that distinguishes those first person tweets from others.

To build a classifier, we generate a training set through crowdsourcing, where workers on Crowdflower\footnote{\url{http://crowdflower.com}} are asked to label whether a tweet is first person. For each topic, we sample 1,000 tweets using proportionate stratification. We group tweets by a user's gender (e.g., male, female, or not-inferred), a user's number of tweets (e.g., the number of tweets$<$100, $>=$100 \& $<$1000, and $>$1000), and user's number of followers (e.g., the number of followers $<$100, $>=$100 \& $<$1000, and $>$1000). Then we sample tweets in which the sample size of each of the group is proportionate to the population size of the same group. Crowd workers then code each tweet according to whether it mentions real activity of an individual (whether one got flu or got fired). Though we call these tweets ``first person'' tweets, they could also be of the kind ``My brother got fired'', mentioning \emph{another} individual.

Firstly, 4.4\% of flu tweets and 19.3\% of unemployment tweets are classified as ``Not related to the topic'' (e.g, spam or out of context). Then, among those on-topic tweets, we find that the unemployment dataset has a higher ratio of reporting real activity: 45\% flu tweets (438/970) and 79\% unemployment tweets (642/812) are classified as a first person tweet.

One difference across the two datasets is that for flu, women tend to correspond to first person tweets, while for unemployment it is men. When looking at the Twitter characteristics of different user groups, users with first person tweets tend to 1) have more favorites; 2) have more tweets; and 3) be less listed than those whose tweets are not first person tweets. We did not find any differences in reporting everyday activity corresponding in number of friends, followers, or days since joining.
% to differences of the number of friends, followers, or days since they join. 

\begin{table}[h!]
\caption{\textbf{Evaluation of classification}}
    \begin{center}
    % \small \frenchspacing
    \begin{tabular}{c|cccc}
    \hline
    \hline
   Topic & Accuracy (\%) & F1 (\%) & Precision (\%) & Recall (\%)\\
    \hline
    Flu & 81 & 84 & 87 & 82\\
    Unemployment & 78 & 83 & 85 & 82 \\
    \hline
    \hline
    \end{tabular}
    \end{center}
    \label{tab:evaluation}
\end{table}

With the training set, we build a first person tweet classifier for each topic using a Random Forest. 
As features we use both lexical features and Twitter profile features. Concretely, we extract n-grams (where n is 1 to 4) from the tweet text and then remove stopwords and words that appear only once. This results in 2,460 words for flu and 2,374 words for unemployment, each of which is used as one feature. As Twitter profile features, we use the number of followers, the number of friends, the number of times listed, the number of favorites, the number of tweets, and the number of days since joining. Thus in total 2,466 features for the flu dataset and 2,380 features for the unemployment dataset were used. 

The classifier was trained on a balanced binary class distribution. We report the error rate estimated using an out-of-bag approach for the random forest bootstraps sample from the balanced data, but re-adjusted to reflect the unbalanced data of the full dataset. For both datasets; the classifier can pick out first person tweets with fairly low error rates (18.86\% for flu and 21.81\% for unemployment) (Table~\ref{tab:evaluation}).

\begin{figure} [h!]
 \begin{center}
    \subfigure[First person users]{\includegraphics[width=.45\textwidth]{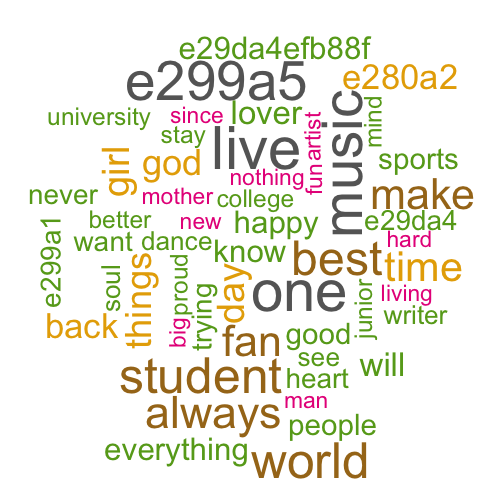}\label{userbio-tagcloud-flu-Firstperson_tweet}}
    \subfigure[Non-first person users]{\includegraphics[width=.45\textwidth]{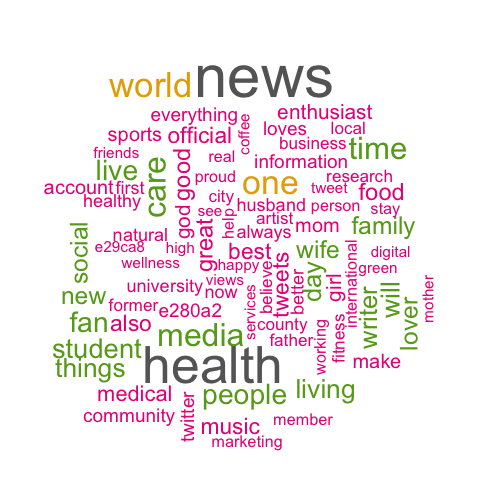}\label{userbio-tagcloud-flu-Non-firstperson_tweet}}
 \label{fig:word_cloud_bio_flu}
 \caption{\textbf{Word cloud for the bios of users tweeting about flu. Left: First person users; right: non-first person. The non-readable tokens refer to unicode emoticons (e.g., e299a5 is a ``heart'' emoticon).}}
 \end{center}
\end{figure}

Inspecting the word cloud (not shown here) for the flu-related tweets that were classified as first person or not, we find that the classifier captures false alarm terms such as `ebola' in the not first person group. However, a difference in the corresponding user sets also emerges when looking at the tag clouds from their bios (Figure~\ref{fig:word_cloud_bio_flu}). Whereas the first person cloud on the left represents more ``normal'' Twitter users, the one on the right hints at more topic oriented (``health'') and news related (``news'') accounts. 

We note that in Figure~\ref{fig:word_cloud_bio_flu}, the non-readable tokens (e.g., ``e299a5'', ``e280a2'') in the wordcloud are unicode codes of emojis. The emoticons appearing in the first person cloud are very positive, either a ``heart'' or ``smiling face'' symbol. For example, ``e299a5'', ``e299a1'', and  ``e29da4'' are all ``heart'' emoticons in different colors and shapes and ``efb88f'' is a smiling face. ``e280a2'' appears in both clouds and represents a ``middle dot'' which is often used to separate words (e.g., enthusiast \textbullet new york) in Twitter.
% ``e29ca8'' is sparkles appear in non-first person cloud. 

\subsection{Demographic Inference}
\label{sec:inference}

For some of our research hypotheses (\textit{[H1], [H4], and [H5]}) we require basic demographic information for the Twitter users.

\noindent \textbf{\emph{Gender:}}
To infer gender, we applied a standard method based on a user's provided first name, using the same name dictionary as in~\cite{magnoweber14socinfo}. Manual inspection showed that this simple approach generally has high precision, at the potential expense of recall.

\noindent \textbf{\emph{Age:}} Due to the lack of longitudinal data for the users in our dataset, we decided not to use a content-based classification approach but, rather, use the \emph{profile picture}. Each profile picture, where present, was passed through the Face++ API\footnote{\url{http://www.faceplusplus.com/demo-detect/}}. When a face is detected, this API returns various bits of information, including an age estimate. Though this image-based inference is undoubtedly noisy, manual inspection for about a dozen personal friends with known age showed that it was by and large surprisingly accurate.

\noindent \textbf{\emph{Geography:}}
To infer state-level location of a Twitter user, we build an algorithm to map location strings to U.S.\ cities and states. The algorithm considers only the locations that mention the country as the U.S.\ or do not mention any country at all, and uses a set of rules to reduce incorrect mappings. When a tweet is geotagged, we map the coordinates to state using a Python library (geopy). If this is not the case, then we look at the location field in their profile. If a state is mentioned, we consider it as their home state. Otherwise, using a dictionary of city to state mapping used in \cite{ChenWO14}, we map city names to states. We call those users whose extracted state is mapped to one of the 51 U.S.\ states (including the federal district Washington, D.C.) ``state-inferred'' users.

\noindent \textbf{\emph{Geographical penetration rate: }}
Though not a demographic variable, we also approximated a state's Twitter penetration rate. To do so, we needed an estimate of where Twitter has more or fewer users. Using the topic-specific datasets would have been inappropriate as, say, one U.S.\ state might be more affected than others by unemployment. Therefore, we used a month of Decahose data (October 2014) to collect tweets for the general terms ``love'', ``like'', ``music'', ``weather'' and ``thing''. The resulting 2,543,219 tweets were mapped to U.S.\ states using the approach described above.
The Twitter user counts for the states in our baseline data were then divided by offline population estimates\footnote{\url{http://www.enchantedlearning.com/usa/states/population.shtml}}. Note that only the \emph{relative} penetration rates are of relevance for our analysis and so collecting more data would likely not affect our results.

\begin{figure} [h!]
 \begin{center}
    \subfigure[The number of patients with flu symptoms (Top) and Flu Tweets (Bottom)]{\includegraphics[width=.45\textwidth]{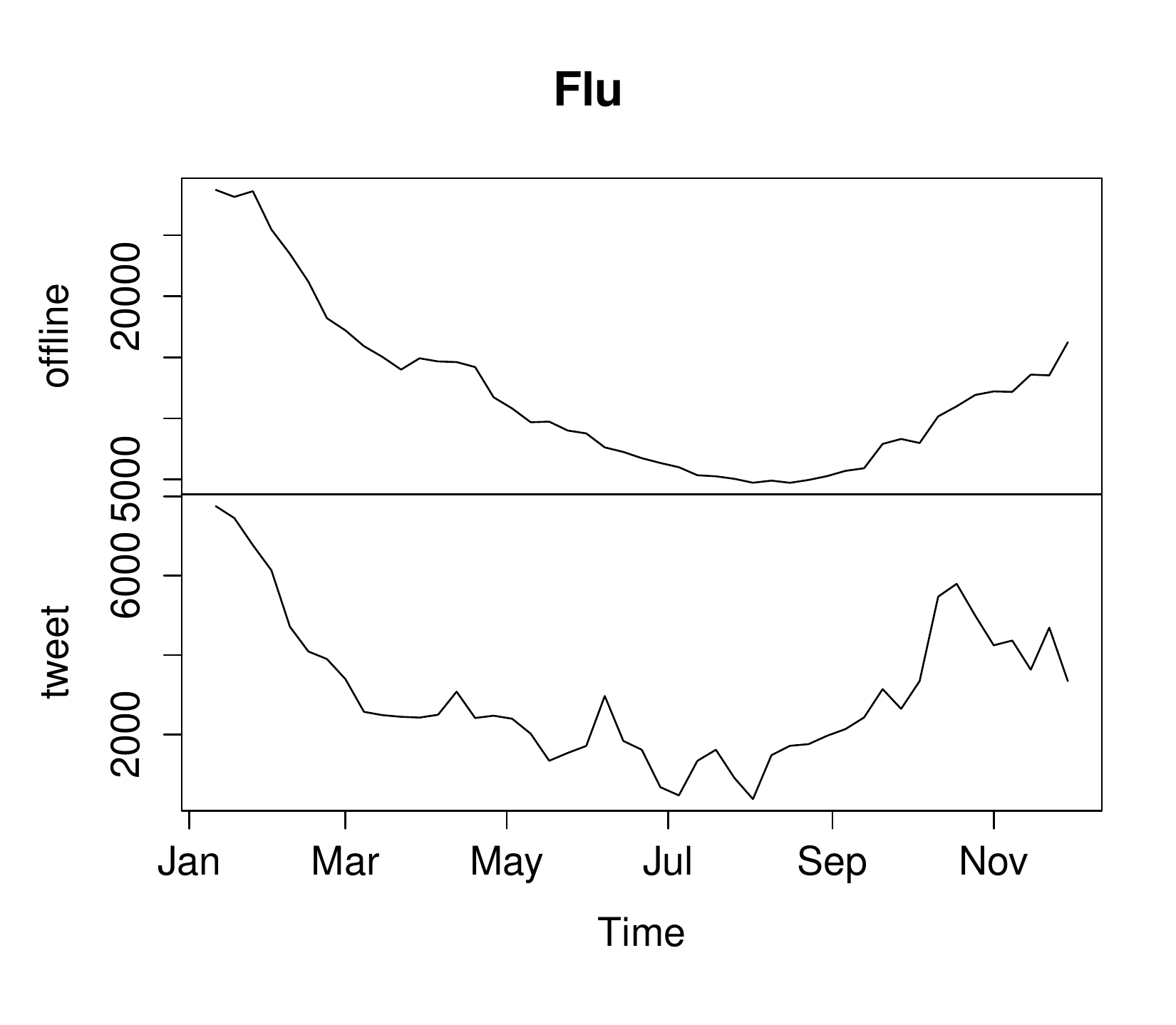}\label{timeseries_flu}}
    \subfigure[Initial Claims for Unemployment Insurance and Job Loss (Top) and Unemployment Tweets (Bottom)]{\includegraphics[width=.45\textwidth]{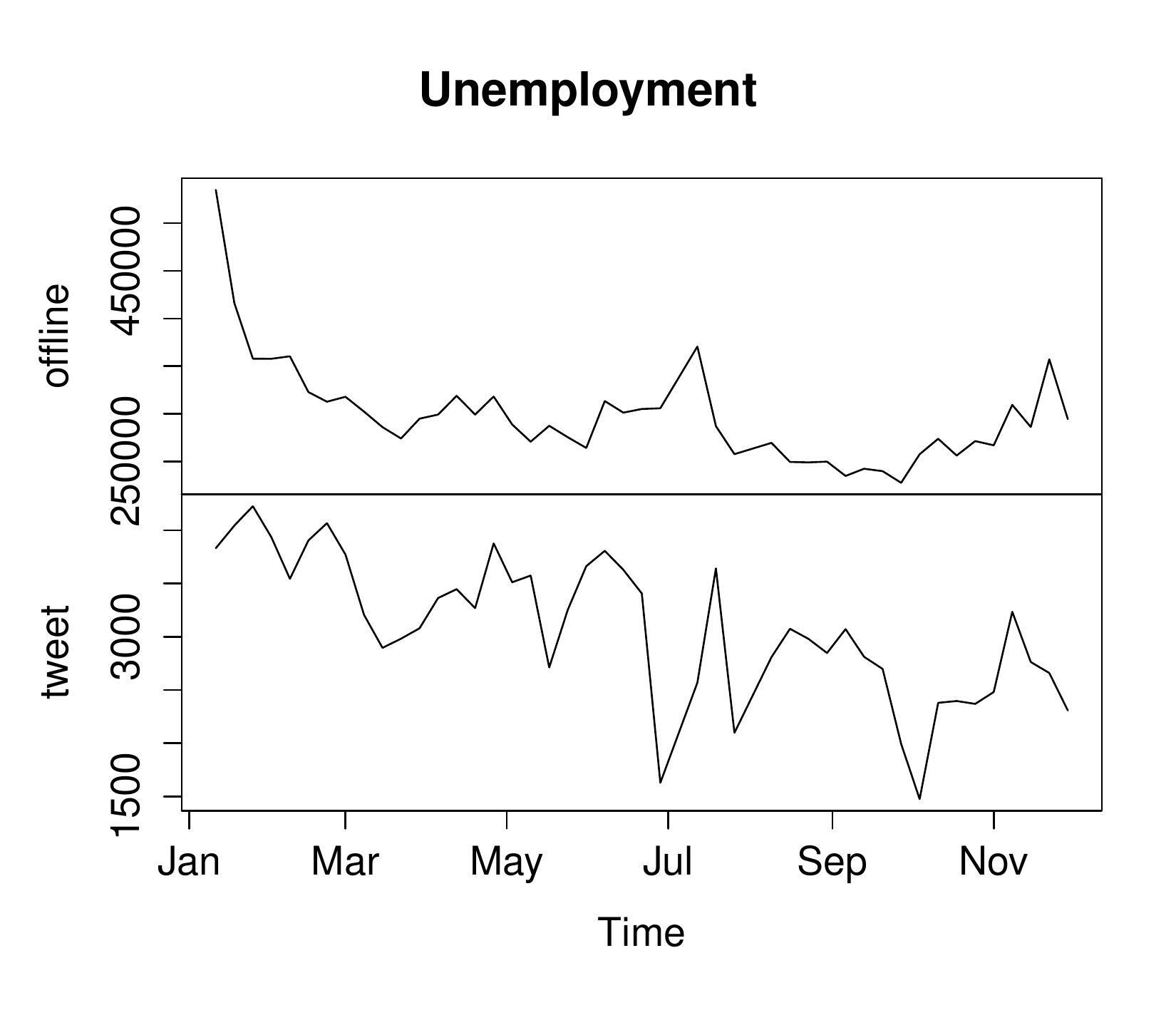}\label{timeseries_unemployment}}
 \label{fig:time_series}
 \caption{\textbf{Time series for flu (left) and unemployment (right). Each plot depicts corresponding offline values and the number of tweets over time. The numbers of tweets are aggregated weekly.}}
 \end{center}
\end{figure}

\section{Time series prediction}

Our time series of flu activity and unemployment rate are weekly counts, i.e., the number of patients with flu symptoms in the week $t$ or the number of people who claimed UI. Likewise, we also accumulated Twitter data weekly as the number of distinct users mentioning related keywords in week $t$. We count the number of unique users per week, not the number of tweets, as this gave more robust results and naturally guarded against simple spam. We will later test our hypotheses by using different user sets and analyzing changes in the now-casting performance. 

Figure~\ref{fig:time_series} shows the time series of offline values (top) and Twitter values (bottom) for flu (left) and unemployment (right) from January to November 2014. Visually, we observe that flu activity data has `smooth' changes over time and that the Twitter data also follows a similar pattern. Correspondingly, the flu dataset shows a high positive Spearman Rank Correlation between offline and Twitter data ($\rho$=0.88, $p<0.005$). The unemployment dataset has several peaks in its offline data and its Twitter data also has more fluctuations. And, yet, it also shows a positive correlation with $\rho$=0.54 ($p<0.005$).

\noindent \textbf{Model}. Given that both of our datasets have positive rank correlations for offline and Twitter data, now-casting seems feasible and we proceed to build a prediction model. We use a simple linear model that uses $n$ lagged values of Twitter time series, which is defined as follows: 
\begin{equation}
{Y_t} = \alpha + \sum_{i=0}^{n} {\gamma_{i} X_{t-i}} + \varepsilon_{t}
\end{equation} 
Here $X_t$ is the number of Twitter users mentioning any keywords of a topic, such as ``flu'', at a given time $t$. The model uses $n$ lagged values of Twitter time series $X_{t-1}$, ... $X_{t-n}$ and the $\gamma_{i}$ are fit to minimize the prediction error.

To test our research hypotheses, we train a model (or a set of models) for a particular user set and then analyze the performance difference on a test set. As a baseline reference model we used data for all users who passed the spam removal process and where their tweets passed the first person classifier. As shown later in this section, the gain from the first person classifier is high, which is why we use this ``First person only'' model as our baseline reference model. Note that we are not trying to propose a particular ``optimal'' prediction model. Rather, we want to assess the effects of user selection strategies on the prediction accuracy. 

\noindent \textbf{Training and testing}. Our offline data is aggregated weekly and our datasets are collected for eleven months (we have 47 data points for both topics). For the purpose of significance test, we generate different set of training and test data. The shortest  training period is 25 weeks with our model and we use 22 weeks of test cases. Then, by moving the window of the training period in increments of one week, we get 21, 20, ..., 1 week of test cases. Then, by extending the window of training period (26, 27, ... 46 weeks), we generate another 21, 20, ... 1 different test cases. We have 253 test cases by moving the window and 231 test cases by extending the window, resulting in 484 test cases in total.

\noindent \textbf{Measure}. Prediction accuracy is measured in terms of the average Mean Absolute Percentage Error (MAPE) which expresses accuracy as a percentage of the error. Because this number is a percentage, it can be easier to understand than other statistics. For example, if the MAPE is 5 then forecast is off by 5\% on average.

\begin{figure} [h!]
 \begin{center}
    \subfigure[Flu]{\includegraphics[width=.85\textwidth]{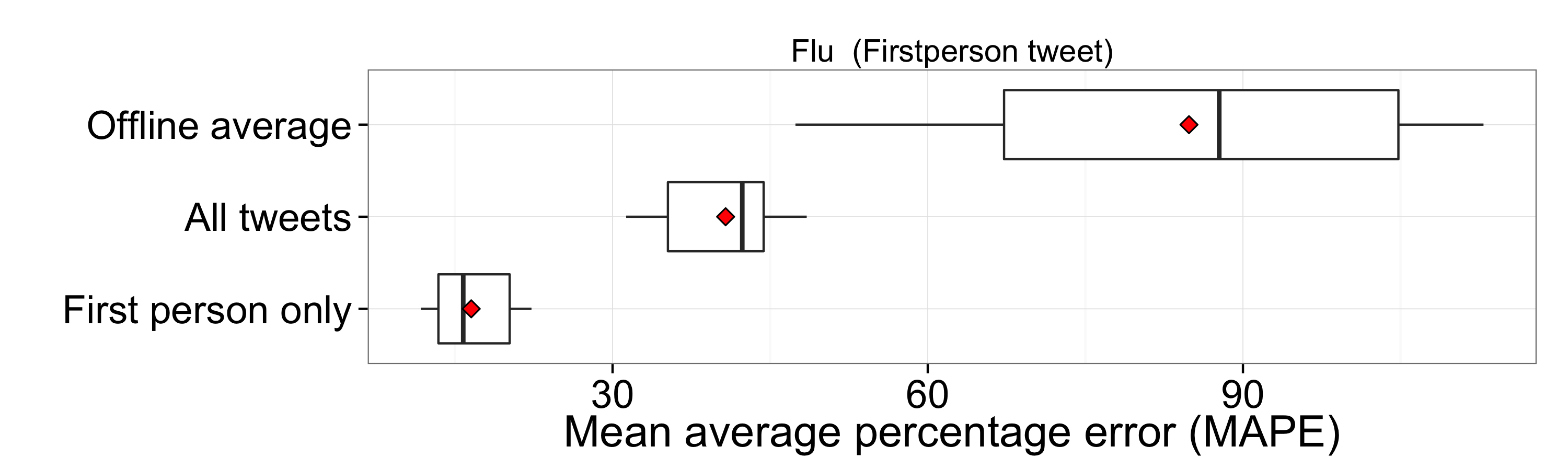}\label{plot_baseline_mape_flu_twitter_only}}
    \subfigure[Unemployment]{\includegraphics[width=.85\textwidth]{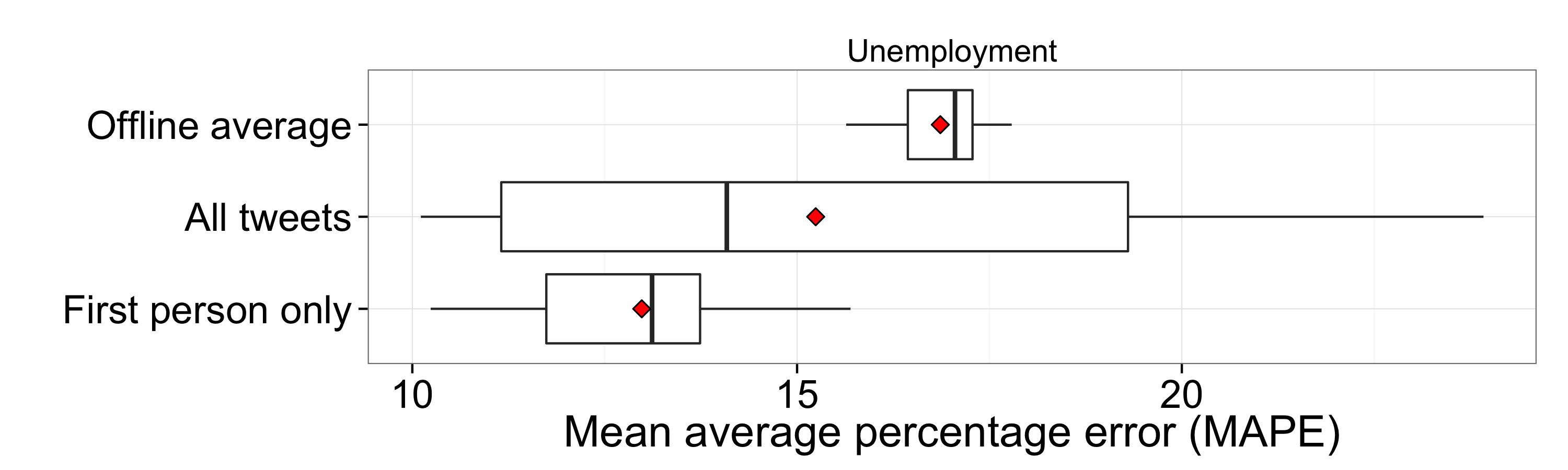}\label{plot_baseline_mape_unemployment_twitter_only}}
 \label{fig:baseline_mape_unemployment}
 \caption{\textbf{MAPE of three basic models. Top: Flu; Bottom: Unemployment.}}
 \end{center}
\end{figure}

\noindent \textbf{Prediction}. The prediction results for three basic models are shown in Figure~\ref{fig:baseline_mape_unemployment}. The box plot for each model shows the distribution of MAPE of our 484 test cases where the bottom of the box is the first quartile, the red dot is the mean, the bar is the median, and the top of the box is the third quartile. 

The top row (``Offline average'') corresponds to a constant prediction of the average of the offline value across 11 months. For example, this model predicts 298,548 cases of reported unemployment throughout all of the test weeks. Note that though the actual prediction is constant, there is still a distribution as the test set is changing. 
The second row (``All tweets'') corresponds to a model that uses a basic ``one tweet, one vote'' approach. 
The final row (``First person only'') is our main baseline corresponding to a model that only considers tweets that passed the first person classifier.

Based on Figure~\ref{fig:baseline_mape_unemployment}, we observe that for both datasets the use of the first person classifier helps, though the improvement is much more pronounced for the flu dataset and only marginal for unemployment. The comparison to the \emph{constant} offline average also shows that, despite its lower MAPE, the unemployment setting is actually harder than the flu setting. As was already evident in Figure~\ref{fig:time_series}, the unemployment time series seem much noisier with a lower inherent correlation. But as the actual fluctuation across the year is small, not even a factor of 2.0 between minimum and maximum, the MAPE appears low.

Note that the constant average incorporates knowledge ``from the future'', i.e., the whole year. As such it could not be used for actual prediction purposes. However, it still helps to shed light on the relative difference in difficulty between the two tasks--it is much harder to gain any improvement for the unemployment setting than for the flu setting.
% unemployment setting is much harder to gain any improvement than the flu setting.

\section{User group and now-casting}
\label{sec:user_group_now_casting}

Having our baseline prediction error rates (using a basic ``tweet, one vote'' approach), we now test our five hypotheses one by one. We note that we only use the first person tweets for testing hypothesis and the ``First person only'' model is our baseline model. The median of MAPE values for the baseline approach (a basic ``one tweet, one vote'' approach) is 15.7 for the flu dataset and 12.7 for the unemployment dataset.

\begin{figure} [h!]
 \begin{center}
    \includegraphics[width=.85\textwidth]{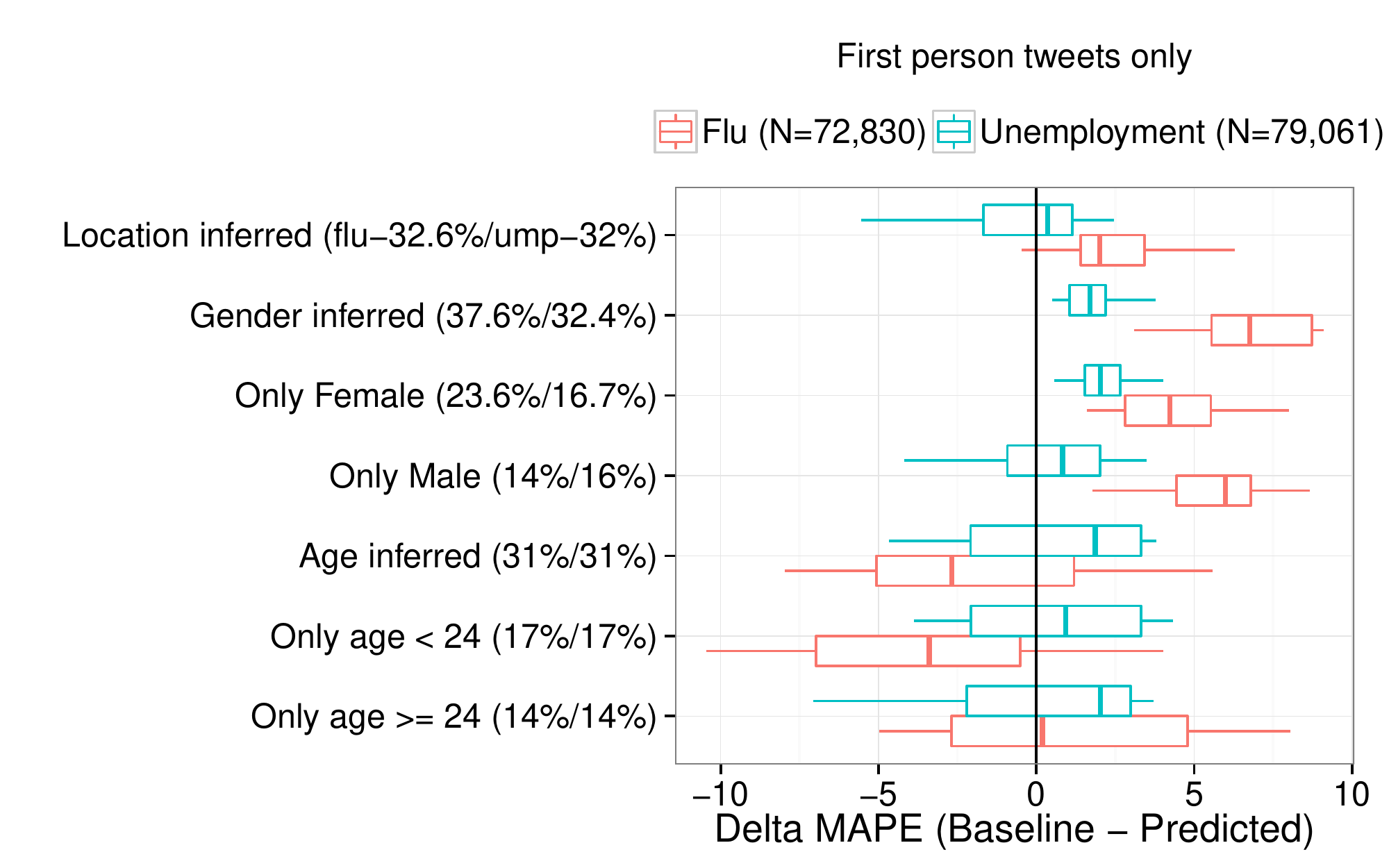}\label{diff_twitter_only_plot_mape_firstperson_userfiltering}
 \label{fig:mape_userfiltering_twitter_only}
 \caption{\textbf{Prediction result of demographic based filtering methods.}}
 \end{center}
\end{figure}

\subsection{More Complete Profile = Better Profile}
\label{sec:complete_profile}

\noindent \textit{[H1] Users with a more complete profile provide better social sensing data.}

We first examine how the prediction result changes when the following three demographic dimensions are used for filtering: 1) requiring a specified location; 2) using the inferred gender; and 3) using the inferred age. 

For each dimension, we create a user set with demographic information inferred. For example we only consider female users (denoted as ``Only females'') and see whether that user group gives a better prediction. Figure~\ref{fig:mape_userfiltering_twitter_only} shows the changes of MAPE, denoted as $\Delta$MAPE (MAPE of our baseline - MAPE of a certain filtering method), for each dataset when using different sets of samples. Note that a positive value indicates improvement over the baseline. In the legend, the number next to the topic is the number of total users considered for the basic approach, and the y-axis label shows the \% of users considered for the corresponding filtering method.

Concerning the use of only gender-inferred users, we find that even though they account for less than 40\% of the total number of users, they give better results for flu activity (median $\Delta$MAPE = $+6.8\%$), and for unemployment claims (median $\Delta$MAPE$=+2\%$). We ran a Mann-Whitney's U test to evaluate the difference in the $\Delta$MAPEs and find a significant effect of gender-based filtering for flu dataset ($U=100, p<0.05$). For the unemployment dataset, the difference was not significant. 

Limiting the users to those with an inferrable gender, based on the name they provide, could serve a range of purposes including additional spam removal or the removal of organizations, and generally yielding more real users. In Section~\ref{sec:babbler} we look more closely at the effect of the fraction of personal accounts on the social sensing performance.

For the flu dataset, the prediction improved again when \emph{further} restricting the user set to only females ($\Delta$MAPE = $+4.1\%$, $U=80, p<0.05$) or to include only males ($\Delta$MAPE = $+7.4\%$, $U=94, p<0.05$). However, when considering only users of estimated age $<$24, the performance dropped slightly ($\Delta$MAPE = $-2.9\%$, $U=19, p<0.05$). A speculative reason could be their exposure to schools and other educational institutes, which could make them less representative of the overall population. 

For the unemployment dataset, none of the methods showed significant differences except the ``Only female'' method ($\Delta$MAPE$=+2.1\%$, $U=76, p<0.05$), though \emph{all} filtering methods led to a (non-significant) improvement in the median, despite using less data than the baseline. 

Filtering methods such as ``Location inferred'', ``Age inferred'', and  ``Only age $>=$ 24'' shows a large variances in prediction result, and thus they were not significant changes.

\begin{figure} [h!]
 \begin{center}
    \includegraphics[width=.85\textwidth]{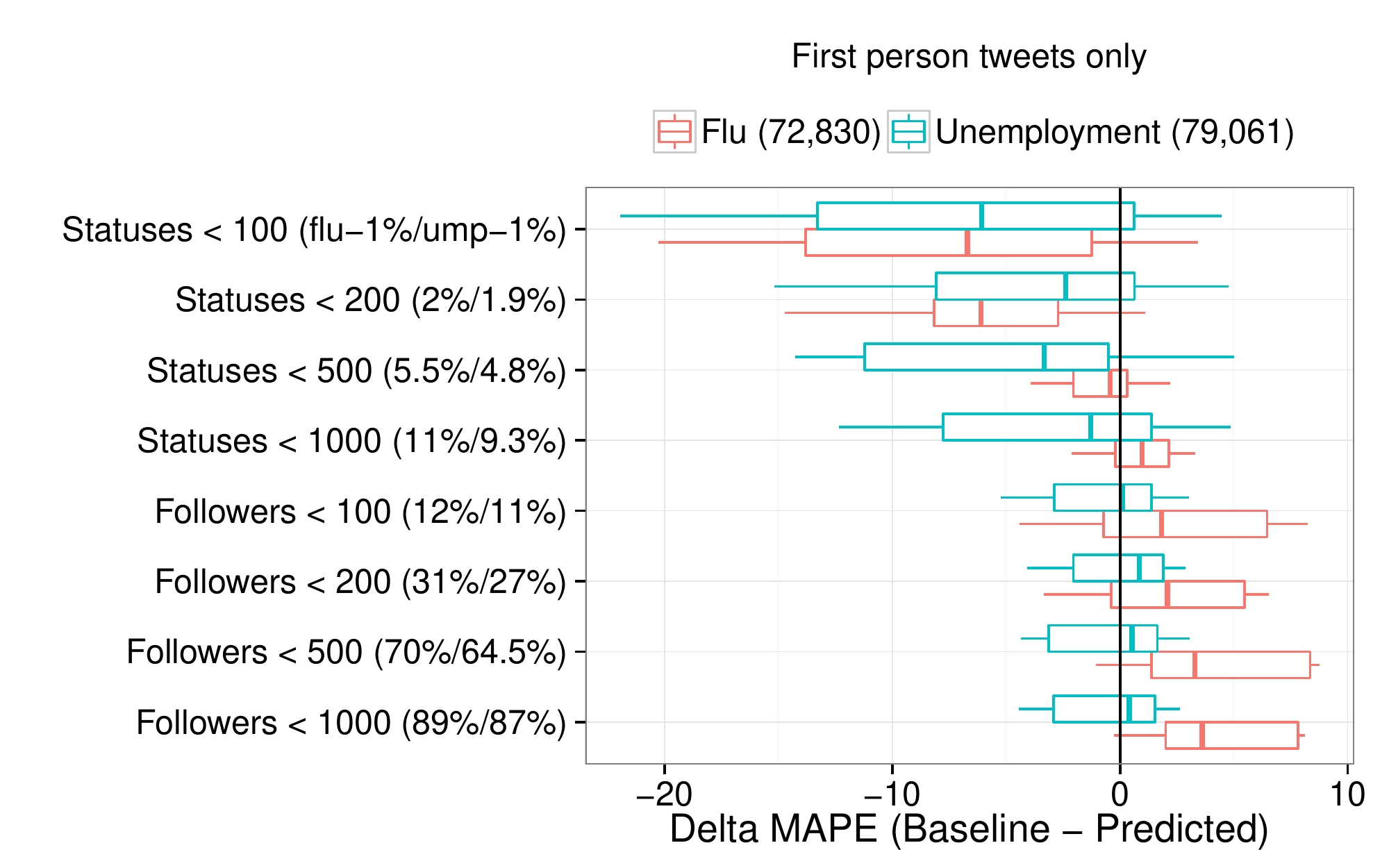}\label{diff_twitter_only_plot_mape_firstperson_hardfiltering}
 \label{fig:mape_hardfiltering_twitter_only}
 \caption{\textbf{Prediction result of Twitter statistics based filtering methods.}}
 \end{center}
\end{figure}

\subsection{Twitter Stats Better Filter Than Demographics} 
\label{sec:twitter_stats}

\noindent \textit{[H2] Twitter statistics, such as a user's number of followers or tweets, are at least as useful as demographic information in determining good social sensors.}

The previous section showed that the mere presence of inferrable demographic information can boost social sensing performance and that, in some cases, further subsetting to a particular demographic subgroup yields additional gains. 
We also experimented with filtering users by Twitter features--the number of followers or the number of tweets. Intuitively, users who have a moderate amount of tweets and followers are ``normal'' users who are better sensors.

Figure~\ref{fig:mape_hardfiltering_twitter_only} shows the results when applying different filtering criteria. 
For the flu dataset, considering users with fewer than 500 followers gives an improvement ($\Delta$MAPE = $+2.6\%$, $U=93, p<0.05$). Users with fewer than 1000 followers increase the performance slightly  ($\Delta$MAPE = $+2.75\%$, $U=93, p<0.05$) with smaller variances.

Filtering users further by the number of followers or by the number of statuses results in less and less remaining data. Later in this section, we will discuss how the data size interacts with the performance (Section~\ref{sec:quailty_vs_quantity}).

\subsection{Babblers are Best}
\label{sec:babbler}

\textit{H3: Users who tweet about their daily lives are better social sensors than those discussing professional or public topics.}

To test the hypothesis, we need to know what fraction of users shares daily activity (e.g., ``I'm off to gym now'') in Twitter. Thus we first classify users based on whether they are individuals or not and how they use Twitter using crowdsourcing. Then, we relate the characteristic of a user group (captured by the distribution of user types it contains) to the now-casting performance of the corresponding user group measure by $\Delta$MAPE.

For the crowdsourced classification, we randomly sample 110 users in each of the following nine user groups (``All'', ``Location inferred'', ``Age inferred'', ``Female'', ``Male'', ``Statuses$<$100'', ``Statuses$<$1000'', ``Followers$<$100'', ``Followers$<$1000''), resulting in 9,961 users for the flu dataset. Since we find that the unemployment dataset does not show meaningful differences of prediction performance across different user groups, we drop it for this task.

Then, crowd workers are provided a link to a Twitter profile, and are asked: 1) whether a Twitter account belongs to an individual, an organization, or is no longer accessible (often indicative of spam identified by Twitter) and 2) whether recent tweets of the Twitter user focus on a single topic or not. When a user is classified as ``individual'', we further ask two additional questions: 1) what is their sub-type--categories are celebrity, is part of an organization, used for personal use, and none of above-- and 2) whether the user shares details about their personal life.

The majority of accounts are labeled ``Individual'' (72\% on average), while ``Organization'' accounts for 6\% of all sampled users. About 22\% were not accessible, meaning that they were active at a point in Jan-Nov 2014, but their page does not exist anymore, often due to suspension by Twitter. For those users who are accessible, 33\% of users are classified as ``Topic-focused''.

For those classified as ``Individual'', 88\% are classified as ``Personal'', leaving us with 11.8\% ``is part of organization'' and 0.2\% of ``Celebrity''. Finally, 71\% of labeled users share their daily life on Twitter. Recall that only users found for certain key words are labeled, so this fraction could be lower for random users.

For each user group, such as ``location inferred'', we compute the following four values: the fraction of organization accounts in a group (denoted as ``Organization''), the fraction of personal accounts in a group (denoted as ``Personal''), the fraction of people focused on a single topic (denoted as ``Topic.focused''), and the fraction of people sharing daily life (denoted as ``Daily.life'').  Using these four values, we wanted to ``explain'' the different performance results for the different user groups. 
As the two variables Personal and Daily.life are highly correlated (Pearson correlation coefficient $r$=0.67 ($p<0.005$)) we include only Personal as an independent variable in a model. 
We then run a linear regression that predicts the now-casting performance ($\Delta~MAPE$) based on a group's characteristics: $\Delta MAPE = \alpha + \beta_1 Personal + \beta_2 Organization + \beta_3 Topic.focused$

\begin{table}[h!] \centering 
  \caption{\textbf{Regression Results for Flu activity}}
  \label{tab:regresion_user_type_flu} 
\begin{tabular}{@{\extracolsep{0pt}}lD{.}{.}{-3} } 
\\[-1.8ex]\hline 
\hline \\[-1.8ex] 
 & \multicolumn{1}{c}{\textit{Dependent variable:}} \\ 
\cline{2-2} 
\\[-1.8ex] & \multicolumn{1}{c}{MAPE} \\ 
\hline \\[-1.8ex] 
 Personal & 0.514^{***} \\ 
  & (0.089) \\ 
  Organization & -0.373^{***} \\ 
  & (0.066) \\ 
  Topic.focused & 0.189^{**} \\ 
  & (0.090) \\ 
  Constant & -47.502^{***} \\ 
  & (9.734) \\ 
 \hline \\[-1.8ex] 
Observations & \multicolumn{1}{c}{152} \\ 
R$^{2}$ & \multicolumn{1}{c}{0.424} \\ 
Adjusted R$^{2}$ & \multicolumn{1}{c}{0.412} \\ 
Residual Std. Error & \multicolumn{1}{c}{4.192 (df = 148)} \\ 
F Statistic & \multicolumn{1}{c}{36.320$^{***}$ (df = 3; 148)} \\ 
\hline 
\hline \\[-1.8ex] 
\textit{Note:}  & \multicolumn{1}{r}{$^{*}$p$<$0.1; $^{**}$p$<$0.05; $^{***}$p$<$0.01} \\ 
\end{tabular} 
\end{table}

Table~\ref{tab:regresion_user_type_flu} shows the regression result. The regression has an adjusted R${^2}$ of 0.412, which means that as much as 41.2\% of the variability of now-casting performance is explained by the combination of three factors. The strongest beta coefficients are registered for Personal ($\beta_1=0.512$), followed by Organization ($\beta_2=-0.373$) and Topic.focused $\beta_3=0.189$) and all coefficients are significant at $p < 0.05$. Though the direction of ``Personal'' (the more the better) and of ``Organization'' (the less the better) are intuitive and in line with our other findings, the sign of the beta coefficient of ``Topic.focused'' is surprising (the more the better). In fact, a linear model built using \emph{only} this factor has the opposite sign, indicating that this coefficient is the result of the correlation with other factors. On its own, the less topic focus the better.

\begin{figure} [h!]
 \begin{center}
    \includegraphics[width=.95\textwidth]{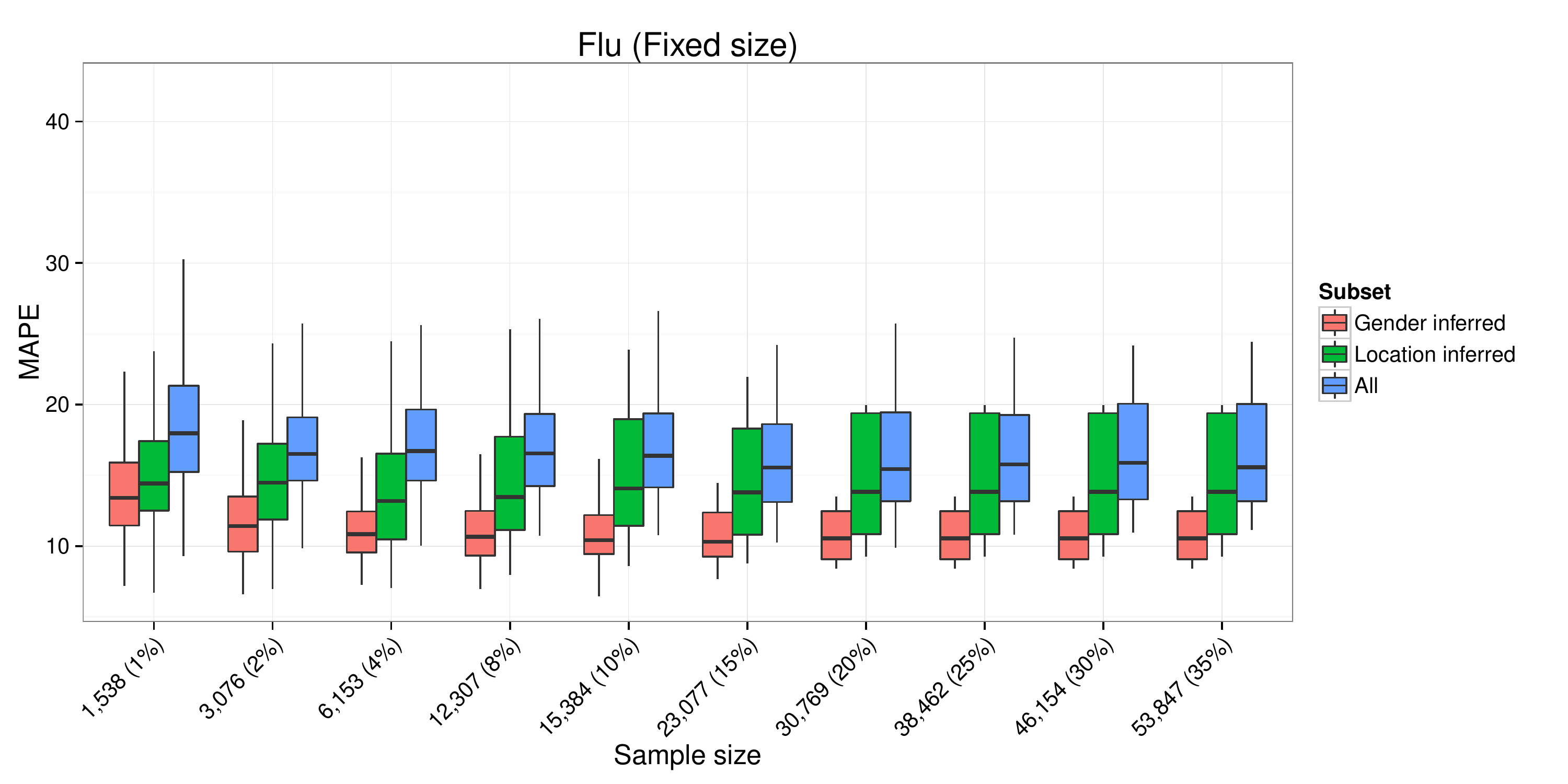}\label{twitter_only_firstperson_plot_by_sample_fixed_size_flu}
 \label{fig:mape_data_quality_vs_quantity_twitter_only}
 \caption{\textbf{Prediction result of three user groups (`All', `Gender-inferred', and `Location-inferred') by different size of dataset.}}
 \end{center}
\end{figure}

\subsection{Better Data Beats Bigger Data}
\label{sec:quailty_vs_quantity}

\textit{[H4] One can obtain equal or better now--casting results by using a drastically reduced dataset, as long as the data quality of the smaller set is high.} 

We have seen that subsetting the data to certain user groups can improve now-casting performance, despite reducing data size. Here we explore the relation between data size and quality in more detail to see how much data we can ``throw away''. 

We randomly sample $N$ tweets ($k$\% of total tweets) where $k$ varies from 1 to 35. Note that the fraction $k$ is with respect to ``All'' tweets and using all available location-inferred tweets is roughly 40\% of all tweets. Then for the following three user groups ``All'', ``Gender-inferred'', and ``Location-inferred'', we aggregate the data into user levels to measure the now-casting performance. For each value $k$, we repeat the experiment 20 times to minimize effects due to sampling variance.

Figure~\ref{fig:mape_data_quality_vs_quantity_twitter_only} shows our results. Generally, the performance degrades slowly for all three user groups, though the confidence intervals become wider. Note that using 10\% of our data, which is itself only 10\% of the Firehose, is in volume equivalent to the 1\% sample available on the Internet Archive\footnote{\url{https://archive.org/details/twitterstream}}. 
Given the general trend, we also believe that having access to the full Firehose with 100\% of public tweets would not greatly benefit the performance.

\begin{figure} [h!]
 \begin{center}
    \includegraphics[width=.85\textwidth]{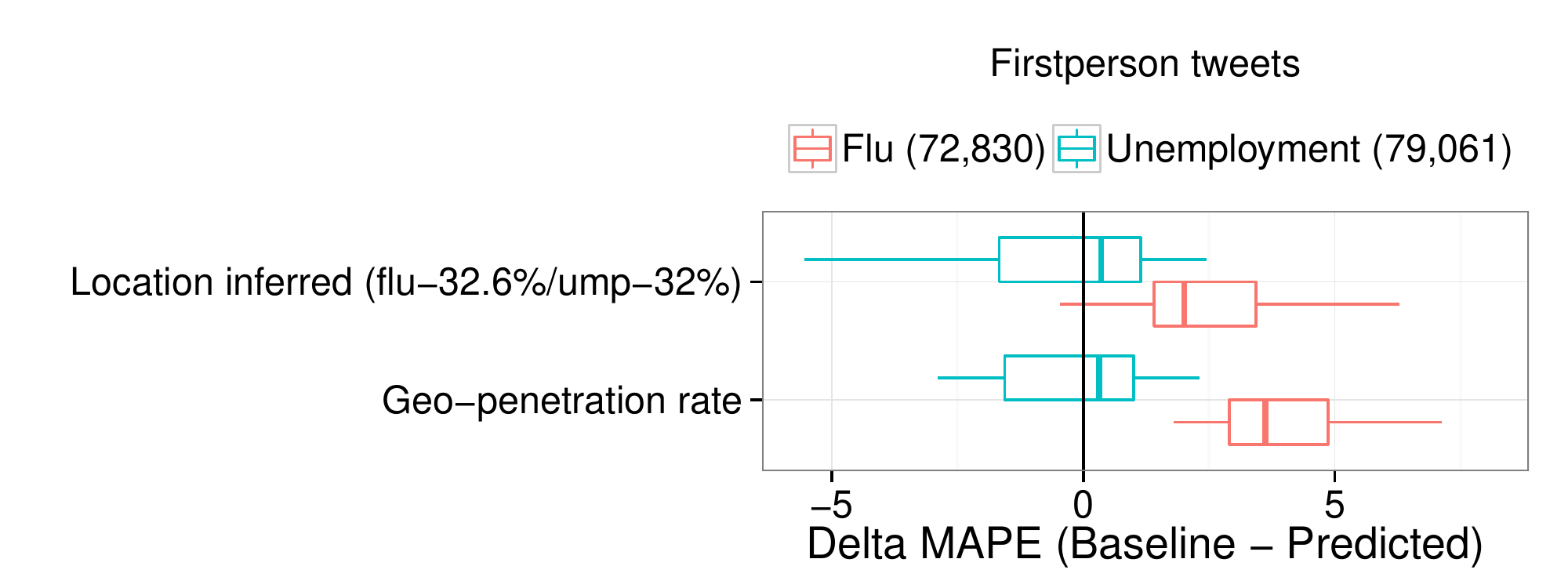}\label{diff_twitter_only_plot_mape_firstperson_penetration}
 \label{fig:mape_penetration_twitter_only}
 \caption{\textbf{Prediction result of geographic reweighting method.}}
 \end{center}
\end{figure}

\subsection{Geographic Reweighting : Putting Weight Where it Belongs} 
\label{sec:geographic_reweighting}

\textit{[H5] By giving more weight to geographical segments of the population that are under--represented on Twitter the now--casting performance increases}.

Internet and Twitter penetration is not uniform across all U.S.\ states. Here, we experiment with a geographic reweighting scheme and whether there are achievable gains. 
Using a linear model, we estimate the number of weekly insurance claims or flu patients of each state using Twitter counts then aggregate them all as follows: $estimated = \sum_{s in U.S.}{nusers_s * \frac{1}{geo_{penetration, s}}}$, where $nusers_s$ is the number of (state-inferred) users mentioning unemployment or flu related keywords in state $s$ for a given week and $geo_{penetration, s}$ is the geographic penetration rate of state $s$ (see Section~\ref{sec:inference}). This weighted sum is then used to fit a time series model as before.

Giving more weight to states with lower Twitter penetration rates improves the prediction result for both domains (Figure~\ref{fig:mape_penetration_twitter_only}) with a 7.1\% improvement of MAPE ($U=79, p<0.05$) for the flu dataset, and a marginal improvement for the unemployment dataset ($\Delta$MAPE = $+0.3\%$, $U=22, p<0.05$). We note that while the ``Location-inferred'' method, which uses exactly the same user set, seems to improve the prediction, it did not pass the significance test.

\begin{figure} [h!]
 \begin{center}
    \includegraphics[width=.85\textwidth]{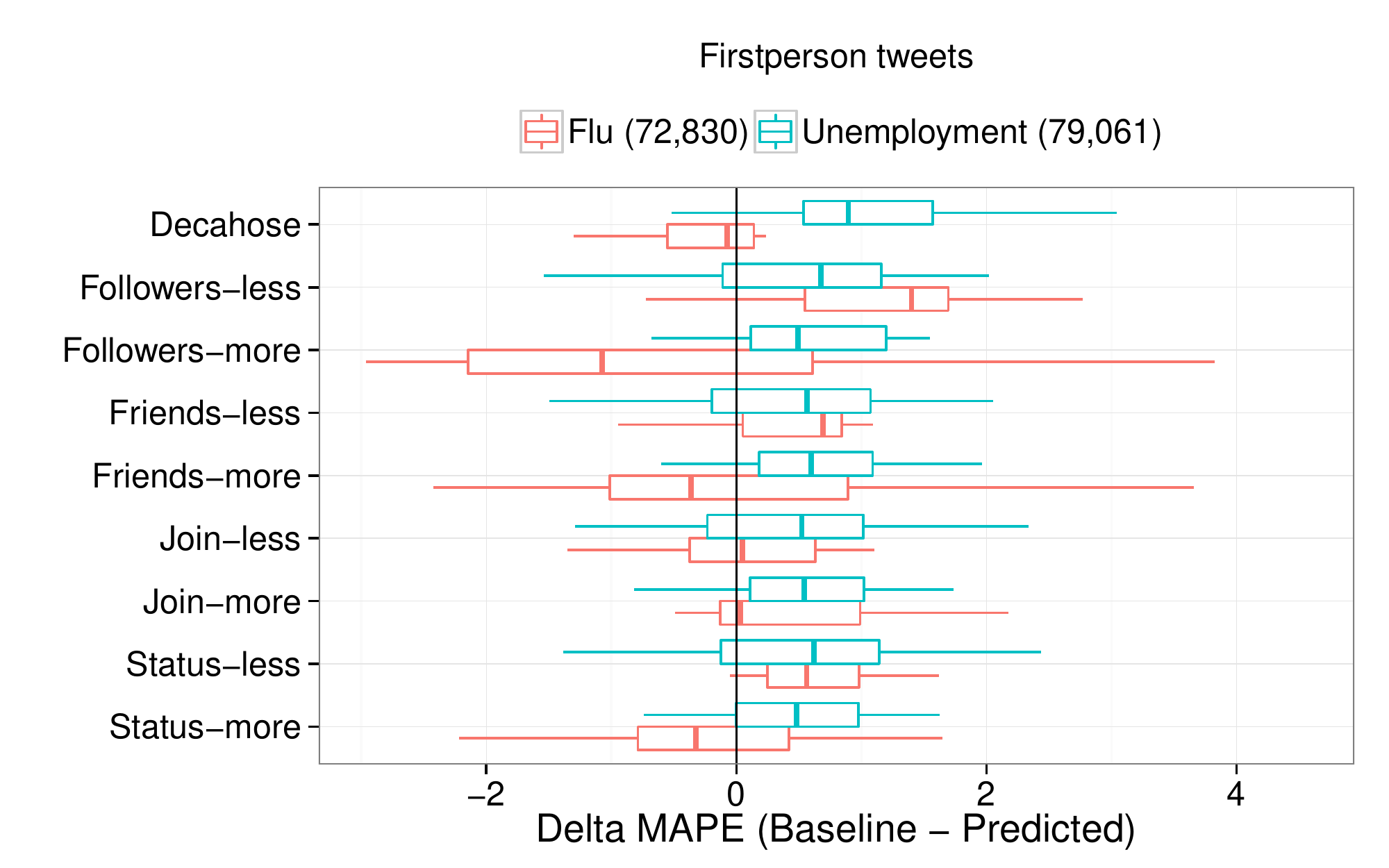}\label{diff_twitter_only_plot_mape_firstperson_silentmajority}
 \label{fig:mape_silentmajority_twitter_only}
 \caption{\textbf{Prediction result of reweighting inactive user method.}}
 \end{center}
\end{figure}

\subsection{Up-Weighting Inactive Users: Boosting the Silent Majority}
\label{sec:silent_majority}

\textit{[H6] By giving more weight to inactive users  the now--casting performance increases}.

One common pitfall of using social media for public opinion monitoring is that the vast majority of users do no express their thoughts on issues such as politics. This also means that, e.g., inferring the political leaning of random users, not just those discussing politics, is harder than one might expect~\cite{cohen2013} and the silent majority is typically ignored. 
Of course it is very difficult to infer whether a person has the flu if the person never tweets their health status. However, for users who do tweet their health status, it makes sense to consider giving different weights to users who tweet constantly and to those that tweet rarely.

To explore ways to weight inactive users, we consider four Twitter features: the number of tweets (status count), the number of followers, the number of friends, and the count of days since the user joined Twitter. 
For each of these four variables we test whether giving more or less weight to large values affects the social sensing performance. 

To assign the weight we used a simple scheme reminiscent of the ``inverse document frequency'' used in information retrieval. Concretely, we give less weight to large values (= active users) and hence, in comparison, more weight to inactive users according to the following function: $ w_{less,u} = \frac{1}{\log_{10}(10 + count_u)}$, where $\mbox{count}_u$ is the corresponding count (e.g., the number of tweets) of user $u$. We also experimented with the opposite scheme where we give more weight to active users: $w_{more,u} = {\log_{10}(10 + count_u)}.$ Both schemes are applied to all four Twitter variables.

We find that only one Twitter feature works: the number of followers. Figure~\ref{fig:mape_silentmajority_twitter_only} shows that, for the flu dataset, the ``less'' scheme, which gives more weight to users who have fewer followers, works better ($\Delta$MAPE = $+1.4\%$, $U=80, p<0.05$) than the ``more'' scheme, which in fact, lowers prediction performance ($\Delta$MAPE = $-1.08\%$, $U=80, p<0.05$). For the unemployment dataset, none of the methods show significant differences.  Given that the number of tweets had no significant effect, the improvement for users with fewer followers could again be a sign that having more ``normal'' users is better.

\section{Offline data in the baseline}
\label{sec:offline_data_in_the_baseline}

So far all of our analysis on social sensing and now-casting has only included online Twitter data as a source for the prediction. Though this is frequently done in research papers, where ``some correlation'' between Twitter and offline indices is shown, it might not be adequate in practice. For example, given the smooth and periodic behavior of the annual flu cases (see Figure~\ref{fig:time_series}) even a prediction using only offline data seems feasible. Furthermore, the offline data could be \emph{combined} with the Twitter data for an improved model. In this section, we compare the Twitter prediction performance to an offline-only model, and we extend the time series model to incorporate both data sources.

\begin{figure} [h!]
 \begin{center}
    \subfigure[Flu]{\includegraphics[width=.85\textwidth]{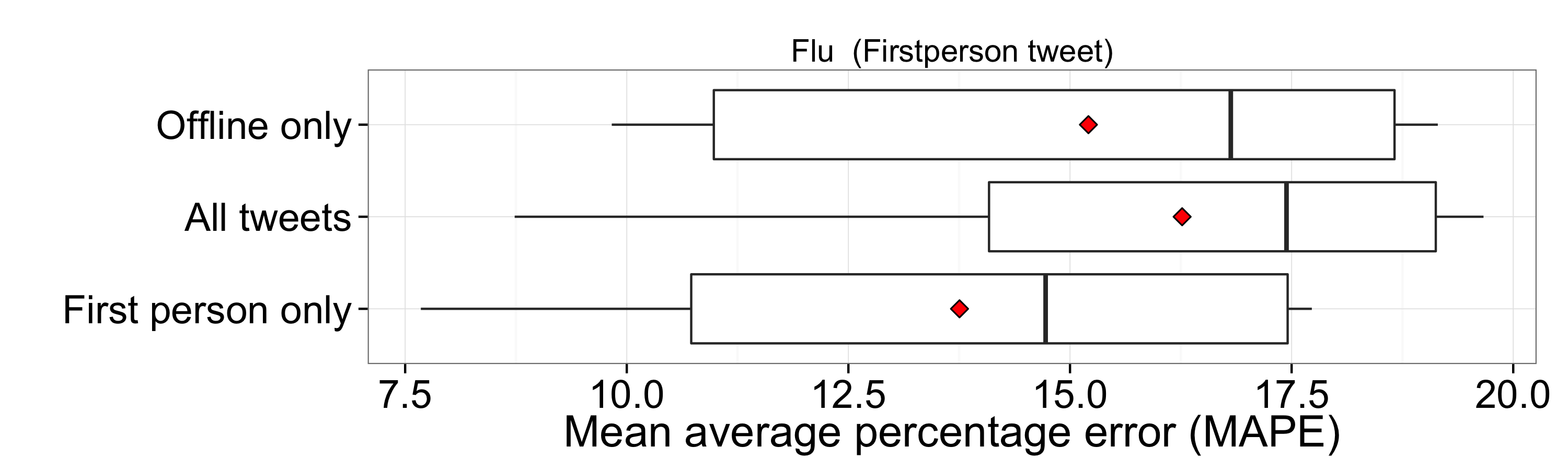}\label{plot_baseline_mape_flu_new}}
    \subfigure[Unemployment]{\includegraphics[width=.85\textwidth]{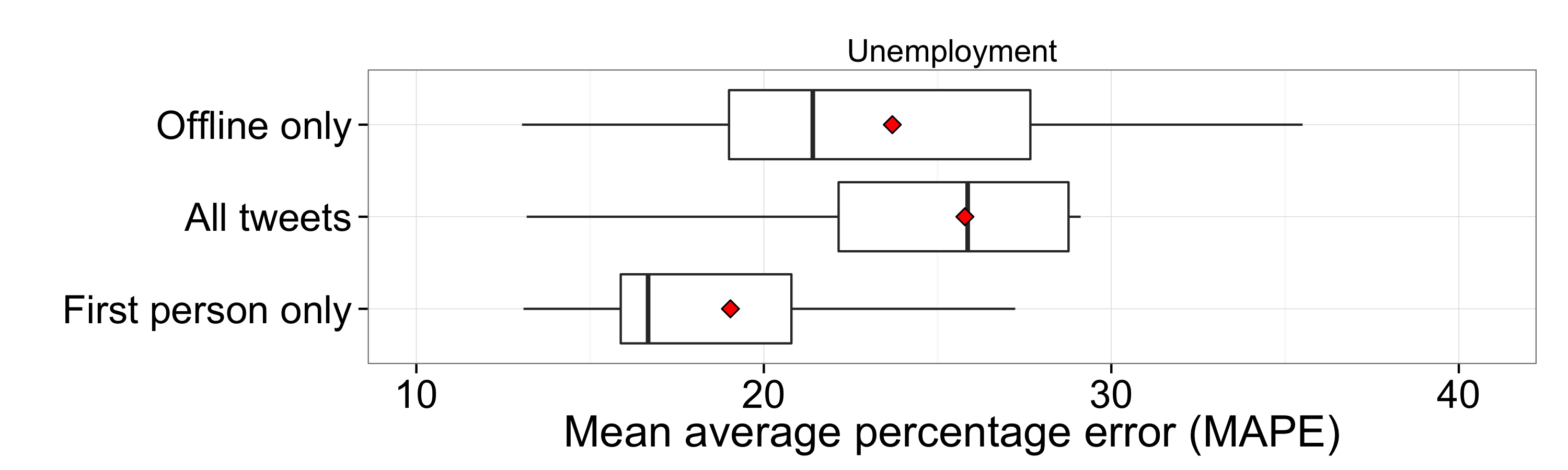}\label{plot_baseline_mape_unemployment_new}}
 \label{fig:baseline_mape_with_offline}
 \caption{\textbf{MAPE of three models including offline data. Top: flu; bottom: unemployment.}}
 \end{center}
\end{figure}

We deploy an Autoregressive Distributed Lag (ARDL) model for predicting current values of flu activity and unemployment rate, incorporating both offline and Twitter values. The ARDL model is defined as follows: 
\begin{equation}
{Y_t} = \alpha + \sum_{i=k}^{m} {\beta_{i} Y_{t-i}} + \sum_{i=0}^{n} {\gamma_{i} X_{t-i}} + \varepsilon_{t},
\end{equation}
where $Y_t$ is the offline value at a given time $t$, and $X_t$ is the number of Twitter users mentioning a related keywords at time $t$. The model uses $m$ lagged values of $Y_t$ (i.e., $Y_{t-k}$, $Y_{t-k-1}$ ... $Y_{t-k-m+1}$) and $n$ lagged values of Twitter time series $X_t$ (i.e., $X_t$, $X_{t-1}$, ... $X_{t-n}$). Note that in our setting the $Y_{t-i}$ values used are shifted by an offset $k$, representing a reporting delay. This is done as the offline values are in practice not available immediately but only with a certain delay -- which is one of the key motivations to use online data in the first place. When combining offline and Twitter data to train a model, we use $k=2, m=2, n=1$ for flu and $k=4, m=2, n=3$ for unemployment. For the offline-only baseline we use the same k, m, and n. We choose $k$ for each domains to be practically realistic--offline data often has a delay in updating survey results. We find that flu activity has a 2 week delay and unemployment has a 4 week delay. Then, using Twitter data, we fill this gap. Given fixed $k$, the $m$ and $n$ were chosen to minimize the error.

Figure~\ref{fig:baseline_mape_with_offline} shows the results for the new offline only baseline, as well as for two models combining (i) all tweets (after spam removal) and (ii) first person tweets with the offline data in an ARDL model. For flu, no statistically significant gains were obtained by using Twitter data when compared to the offline only baseline. For unemployment, use of offline data \emph{hurts} the predictive performance, most likely due to the large amounts of inherent noise in the time series (see Figure~\ref{fig:time_series}). 
In the following section, we will discuss the implications of these findings for social sensing in general.

\section{Discussion: The Meaning of it All}
\label{sec:discussion}

Using social media for now-casting is often motivated by a ``social sensor'' analogy. Usually, it is implicitly assumed that if a user mentions something then this is because they are affected by it. For example, they might have the flu or they might have lost their job and are providing an ``honest signal''\footnote{Note that we use the term ``honest signal'' without any reference to the term's meaning in evolutionary biology.}.

In this paper we set out to shed light on how social media sensing \emph{actually} works by looking at what types of users are the best social sensors. For example, it is not a priori clear that monitoring only curated news references to the flu (a.k.a., high quality data) would lead to worse performance than monitoring individuals' noisy tweets. However, by and large, we do find that now-casting performance is at its best when the fraction of ``normal'' people is at its highest. Even discarding those high quality (and thus truthful and credible) tweets helped to improve now-casting performance, one important ingredient in ``normal'' is the identification of actual first person tweets (see Section~\ref{sec:firstperson_tweet_classifier}), which helps provide a cleaner signal. Furthermore, ``normal'' is linked to having a complete profile (H1), not having too many followers (H2), and generally ``babbling'' about personal life (H3).

In most papers describing the use of social media for now-casting, lagged offline data is not incorporated in the prediction models and, worse, often not even included in a baseline. In Section~\ref{sec:offline_data_in_the_baseline} we showed that offline data should generally be integrated into the model if the ultimate goal is really to have the most accurate prediction performance possible. However, in this paper we focused on ``what makes social sensing work'' rather than ``how to get the best possible now-casting performance''. Domain experts looking at a particular task should carefully consider which additional data sources beyond social media they could incorporate in their models and options range from Google Trends data\footnote{\url{http://www.google.com/trends/}} to weather forecasts. Specialized time series models, e.g., including different periodicities, might also be worth considering. Generally, a one-size-fits-all approach to now-casting is unlikely to yield the best performance, and experimenting solely with a Twitter-only dataset is artificially restrictive.

Our improvements with respect to the ``one tweet, one vote'' baseline for first-person-only tweets are arguably small, +0.3\% for unemployment data in Section 6.5 and +7.4\% for flu data in Section 6.1. The main contribution of this paper is, however, methodological and goes beyond improving now-casting performance by a couple of percent. Rather, the emphasis of the paper is on understanding which types of user groups contribute to social sensing performance in general. A priori, it is not obvious if the ``always-on babblers'' provide better signals than more reserved and actively filtering social media users. We believe that the insights obtained regarding this question are likely to apply to other social media such as Instagram as well, whereas methods to improve now-casting for Twitter are likely to be more specific.

One appropriate question to ask is why, compared to a constant ``offline average'' baseline, significant improvements were achieved using Twitter data for the case of flu, but not for unemployment (see Figure~\ref{fig:baseline_mape_unemployment}). 
One potential reason is that the choice of keywords may be poor and lead to too many false positives, though the same set had previously been used in \cite{Culotta13}. However, crowd-labeling showed that more than 79\% tweets are on-topic first person tweets compared to 45\% for flu. Another reason could relate to a difference in the propensity to report on losing a job, which could come with a loss of respect. Similarly, people are likely more hesitant to report on having herpes, compared to having the flu. There could also be data-intrinsic changes, either in how Twitter is used or how unemployment evolves over time, that cause poor performance. We observe that the Twitter-based prediction at \url{http://econprediction.eecs.umich.edu/} also has poor performance for the period of 2014. Finally, while flu and other easily transmittable diseases might only be weakly linked to economic status, this is different for unemployment. The better-off-than-average Twitter population might hence undergo a very different dynamics than the rest of the population.

As a preliminary analysis, we tested how close the first person flu and the first person unemployment Twitter users are to ``normal'' Twitter users. To have a reference set of normal users, we collected a set of users tweeting about general terms such as `music', `weather' or `thing'. Then we extracted the terms from the bios of this reference user set and compared the terms found to those in the bios of the (i) first person flu users and (ii) first person unemployment users. The comparison was done using Kendall Tau rank correlation on the terms sorted by how many users used them. The flu-vs.-reference similarity was higher (.57) than the unemployment-vs.-reference similarity (.43), indicating again that the more normal the users the better they are for social sensing.

Note that in particular domains, other dynamics might be at play.  For instance, if there were enough cinemas tweeting their daily program then monitoring these tweets would give a good indication of the number of movies showing a particular movie which, in turn, is expected to be strongly correlated with the number of people watching it. In this case, ``bots'' are actually great sensors. Similarly, for predicting stock trends monitoring ``experts'' might be more promising than monitoring normal people.

We close the discussion with a set of recommendations that we expect to work in social sensing settings where (i) frequent, reliable ground truth is available, (ii) there is little stigma associated with publicly admitting of being affected, and (iii) where the effects of astro-turfing are expected to be minimal. 

\begin{itemize}
\item A first person classifier helps to improve data quality (Section~\ref{sec:firstperson_tweet_classifier}).
\item Limit the user set to those with a proper user profile, in particular those with a proper name (H1).
\item Crowd-sourced labeling for the fraction of personal accounts can provide indications as to which subset will work best (H3).
\item Surprisingly little data performs well (H4) and so using the historic 1\% sample could help.%\footnote{\url{https://archive.org/details/twitterstream}}
\item Using geographic re-weighting to correct for different penetration rates can help (H5).
\item Experiment with filtering/reweighting users with different Twitter statistics (H2, H6).
\item Include offline data and other sources in the prediction model (Section~\ref{sec:offline_data_in_the_baseline}).
\item Compare your users to a reference set to quantify how ``normal'' they are (see discussion above).
\end{itemize}

\section{Conclusion}
In this paper, we looked at what makes social media social sensing work by looking at which user groups are the best social sensors for now-casting applications. We went beyond the usual ``one tweet, one vote'' approach and experimented with a number of filtering and reweighting schemes. We showed that, in general, ``normal'' users tweeting about their personal lives are the best social sensors. In a dedicated Discussion section we also gave a number of concrete suggestions, including ways to measure how ``normal'' a given user set is.
 
To the best of our knowledge, this is the first study that systematically looks at who should be used as a social sensor - and who should not. We believe that our findings and suggestions are useful for a wide range of social sensing and now-casting tasks.

\bibliographystyle{abbrv}
\bibliography{epj-nowcasting-arxiv}

\begin{thebibliography}{10}

\bibitem{AggarwalA13}
C.~C. Aggarwal and T.~F. Abdelzaher.
\newblock Social sensing.
\newblock In {\em Managing and Mining Sensor Data}, pages 237--297. 2013.

\bibitem{AlbakourMO13}
M.~Albakour, C.~Macdonald, and I.~Ounis.
\newblock Identifying local events by using microblogs as social sensors.
\newblock In {\em OAIR}, pages 173--180, 2013.

\bibitem{AliSSONM11}
R.~Ali, C.~Sol{\'{\i}}s, M.~Salehie, I.~Omoronyia, B.~Nuseibeh, and W.~Maalej.
\newblock Social sensing: when users become monitors.
\newblock In {\em SIGSOFT}, pages 476--479, 2011.

\bibitem{AlowibdiBY13}
J.~S. Alowibdi, U.~A. Buy, and P.~S. Yu.
\newblock Language independent gender classification on twitter.
\newblock In {\em ASONAM}, pages 739--743, 2013.

\bibitem{antenucci2014}
D.~Antenucci, M.~Cafarella, M.~Levenstein, C.~R{\'e}, and M.~D. Shapiro.
\newblock {Using Social Media to Measure Labor Market Flows}.
\newblock Mar. 2014.
\newblock
  \url{http://www-personal.umich.edu/~shapiro/papers/LaborFlowsSocialMedia.pdf}.

\bibitem{AramakiMM11}
E.~Aramaki, S.~Maskawa, and M.~Morita.
\newblock Twitter catches the flu: Detecting influenza epidemics using twitter.
\newblock In {\em EMNLP}, pages 1568--1576, 2011.

\bibitem{bollen2011modeling}
J.~Bollen, H.~Mao, and A.~Pepe.
\newblock Modeling public mood and emotion: Twitter sentiment and
  socio-economic phenomena.
\newblock In {\em ICWSM}, 2011.

\bibitem{bordino2012web}
I.~Bordino, S.~Battiston, G.~Caldarelli, M.~Cristelli, A.~Ukkonen, and
  I.~Weber.
\newblock Web search queries can predict stock market volumes.
\newblock {\em PloS one}, 7(7):e40014, 2012.

\bibitem{Bravo-MarquezGMP12}
F.~Bravo{-}Marquez, D.~Gayo{-}Avello, M.~Mendoza, and B.~Poblete.
\newblock Opinion dynamics of elections in twitter.
\newblock In {\em LA-WEB}, pages 32--39, 2012.

\bibitem{CepniA14}
K.~Cepni and {\"{O}}.~B. Akan.
\newblock Social sensing model and analysis for event detection and estimation
  with twitter.
\newblock In {\em CAMAD}, pages 31--35, 2014.

\bibitem{ChenWO14}
L.~Chen, I.~Weber, and A.~Okulicz{-}Kozaryn.
\newblock {U.S.} religious landscape on twitter.
\newblock In {\em SocInfo}, pages 544--560, 2014.

\bibitem{ChengCL10}
Z.~Cheng, J.~Caverlee, and K.~Lee.
\newblock You are where you tweet: a content-based approach to geo-locating
  twitter users.
\newblock In {\em CIKM}, pages 759--768, 2010.

\bibitem{cohen2013}
R.~Cohen and D.~Ruths.
\newblock {Classifying Political Orientation on Twitter: It's Not Easy!}
\newblock In {\em ICWSM}, 2013.

\bibitem{Culotta13}
A.~Culotta.
\newblock Lightweight methods to estimate influenza rates and alcohol sales
  volume from twitter messages.
\newblock {\em Language Resources and Evaluation}, 47(1):217--238, 2013.

\bibitem{Gayo-Avello12}
D.~Gayo{-}Avello.
\newblock No, you cannot predict elections with twitter.
\newblock {\em {IEEE} Internet Computing}, 16(6):91--94, 2012.

\bibitem{ginsberg2008detecting}
J.~Ginsberg, M.~H. Mohebbi, R.~S. Patel, L.~Brammer, M.~S. Smolinski, and
  L.~Brilliant.
\newblock Detecting influenza epidemics using search engine query data.
\newblock {\em Nature}, 457(7232):1012--1014, 2008.

\bibitem{gonzalez2014assessing}
S.~Gonz{\'a}lez-Bail{\'o}n, N.~Wang, A.~Rivero, J.~Borge-Holthoefer, and
  Y.~Moreno.
\newblock Assessing the bias in samples of large online networks.
\newblock {\em Social Networks}, 38:16--27, 2014.

\bibitem{IkawaET12}
Y.~Ikawa, M.~Enoki, and M.~Tatsubori.
\newblock Location inference using microblog messages.
\newblock In {\em WWW}, pages 687--690, 2012.

\bibitem{DavisPOA11}
C.~A.~D. Jr., G.~L. Pappa, D.~R.~R. de~Oliveira, and F.~de~Lima~Arcanjo.
\newblock Inferring the location of twitter messages based on user
  relationships.
\newblock {\em T. {GIS}}, 15(6):735--751, 2011.

\bibitem{LamposBC10}
V.~Lampos, T.~D. Bie, and N.~Cristianini.
\newblock Flu detector - tracking epidemics on twitter.
\newblock In {\em ECML-PKDD}, pages 599--602, 2010.

\bibitem{lazer2014parable}
D.~M. Lazer, R.~Kennedy, G.~King, and A.~Vespignani.
\newblock The parable of google flu: Traps in big data analysis.
\newblock {\em Science}, 2014.

\bibitem{LiuYLF11}
S.~Liu, J.~Yang, B.~Li, and C.~Fu.
\newblock Volunteer sensing: The new paradigm of social sensing.
\newblock In {\em ICPADS}, pages 982--987, 2011.

\bibitem{liu2013}
W.~Liu and D.~Ruths.
\newblock {What's in a name? Using first names as features for gender inference
  in Twitter}.
\newblock {\em Analyzing Microtext: 2013 AAAI Spring Symposium}, 2013.

\bibitem{liu2014tweets}
Y.~Liu, C.~Kliman-Silver, and A.~Mislove.
\newblock The tweets they are a-changin’: Evolution of twitter users and
  behavior.
\newblock In {\em ICWSM}, 2014.

\bibitem{MadanCLP10}
A.~Madan, M.~Cebri{\'{a}}n, D.~Lazer, and A.~Pentland.
\newblock Social sensing for epidemiological behavior change.
\newblock In {\em UbiComp}, pages 291--300, 2010.

\bibitem{magnoweber14socinfo}
G.~Magno and I.~Weber.
\newblock International gender differences and gaps in online social networks.
\newblock In {\em SocInfo}, pages 121--138, 2014.

\bibitem{MahmudND12}
J.~Mahmud, J.~Nichols, and C.~Drews.
\newblock Where is this tweet from? inferring home locations of twitter users.
\newblock In {\em ICWSM}, 2012.

\bibitem{MetaxasMG11}
P.~T. Metaxas, E.~Mustafaraj, and D.~Gayo{-}Avello.
\newblock How (not) to predict elections.
\newblock In {\em PASSAT/SocialCom}, pages 165--171, 2011.

\bibitem{amislove2011@icwsm}
A.~Mislove, S.~Lehmann, Y.~Y. Ahn, and J.~P. Onnela.
\newblock {Understanding the Demographics of Twitter Users.}
\newblock {\em ICWSM}, 2011.

\bibitem{morstatter2013sample}
F.~Morstatter, J.~Pfeffer, H.~Liu, and K.~M. Carley.
\newblock Is the sample good enough? comparing data from twitter's streaming
  api with twitter's firehose.
\newblock In {\em ICWSM}, 2013.

\bibitem{nguyen2013@icwsm}
D.~Nguyen, R.~Gravel, D.~Trieschnigg, and T.~Meder.
\newblock {" How Old Do You Think I Am?" A Study of Language and Age in
  Twitter.}
\newblock {\em ICWSM}, 2013.

\bibitem{NguyenL14}
M.~Nguyen and E.~Lim.
\newblock On predicting religion labels in microblogging networks.
\newblock In {\em SIGIR}, pages 1211--1214, 2014.

\bibitem{connor2010@icwsm}
B.~O'Connor, R.~Balasubramanyan, B.~R. Routledge, and N.~A. Smith.
\newblock From tweets to polls: Linking text sentiment to public opinion time
  series.
\newblock {\em ICWSM}, 11:122--129, 2010.

\bibitem{OlguinP08}
D.~O. Olgu{\'{\i}}n and A.~Pentland.
\newblock Social sensors for automatic data collection.
\newblock In {\em AMCIS}, page 171, 2008.

\bibitem{PennacchiottiP11}
M.~Pennacchiotti and A.~Popescu.
\newblock Democrats, republicans and starbucks afficionados: user
  classification in twitter.
\newblock In {\em KDD}, pages 430--438, 2011.

\bibitem{PennacchiottiP11b}
M.~Pennacchiotti and A.~Popescu.
\newblock A machine learning approach to twitter user classification.
\newblock In {\em ICWSM}, 2011.

\bibitem{PontesMV0AKA12}
T.~Pontes, G.~Magno, M.~A. Vasconcelos, A.~Gupta, J.~M. Almeida, P.~Kumaraguru,
  and V.~Almeida.
\newblock Beware of what you share: Inferring home location in social networks.
\newblock In {\em ICDM-WS}, pages 571--578, 2012.

\bibitem{preis2012quantifying}
T.~Preis, D.~Y. Kenett, H.~E. Stanley, D.~Helbing, and E.~Ben-Jacob.
\newblock Quantifying the behavior of stock correlations under market stress.
\newblock {\em Scientific reports}, 2, 2012.

\bibitem{SakakiOM10}
T.~Sakaki, M.~Okazaki, and Y.~Matsuo.
\newblock Earthquake shakes twitter users: real-time event detection by social
  sensors.
\newblock In {\em WWW}, pages 851--860, 2010.

\bibitem{SzomszorKQ10}
M.~Szomszor, P.~Kostkova, and E.~de~Quincey.
\newblock {\#}swineflu: Twitter predicts swine flu outbreak in 2009.
\newblock In {\em eHealth}, pages 18--26, 2010.

\bibitem{TumasjanSSW10}
A.~Tumasjan, T.~O. Sprenger, P.~G. Sandner, and I.~M. Welpe.
\newblock Predicting elections with twitter: What 140 characters reveal about
  political sentiment.
\newblock In {\em ICWSM}, 2010.

\bibitem{ZagheniGWS14}
E.~Zagheni, V.~R.~K. Garimella, I.~Weber, and B.~State.
\newblock Inferring international and internal migration patterns from twitter
  data.
\newblock In {\em WWW}, pages 439--444, 2014.

\bibitem{Zagheni2015}
E.~Zagheni and I.~Weber.
\newblock Demographic research with non-representative internet data.
\newblock {\em International Journal of Manpower}, 36(1):13--25, 2015.

\bibitem{ZamalLR12}
F.~A. Zamal, W.~Liu, and D.~Ruths.
\newblock Homophily and latent attribute inference: Inferring latent attributes
  of twitter users from neighbors.
\newblock In {\em ICWSM}, 2012.

\end{thebibliography}

\end{document}